\documentclass[authoryear,12pt]{elsarticle}
\usepackage[T1]{fontenc}
\usepackage[utf8]{inputenc}
\usepackage{tabularx}
\usepackage{siunitx}
\usepackage{color}
\usepackage{colortbl}
\usepackage{pdflscape}
\usepackage{todonotes}
\usepackage{rotating}

\makeatletter
\def\ps@pprintTitle{%
 \let\@oddhead\@empty
 \let\@evenhead\@empty
 \def\@oddfoot{}%
 \let\@evenfoot\@oddfoot}
\makeatother

\usepackage{amssymb}
\usepackage{amsthm}

\usepackage[htt]{hyphenat}
\usepackage[hidelinks]{hyperref}

\journal{}

\makeatletter
\providecommand{\doi}[1]{%
  \begingroup
    \let\bibinfo\@secondoftwo
    \urlstyle{rm}%
    \href{http://dx.doi.org/#1}{%
      doi:\discretionary{}{}{}%
      \nolinkurl{#1}%
    }%
  \endgroup
}
\makeatother

\begin{document}

\begin{frontmatter}



\title{Are Investors Biased Against Women? Analyzing How Gender Affects Startup Funding in Europe} 


\author{Michael Färber}

\address{Karlsruhe Institute of Technology (KIT), Institute AIFB, Karlsruhe, Germany\\michael.faerber@kit.edu}

\author{Alexander Klein}

\address{Karlsruhe Institute of Technology (KIT), Institute AIFB, Karlsruhe, Germany\\x.klein.14@gmail.com}

\begin{abstract}
One of the main challenges of startups is to raise capital from investors. For startup founders, it is therefore crucial to know whether investors have a bias against women as startup founders and in which way startups face disadvantages due to gender bias. Existing works on gender studies have mainly analyzed the US market.
In this paper, we aim to give a more comprehensive picture of gender bias in early-stage startup funding. We examine European startups listed on Crunchbase using Semantic Web technologies and analyze how the share of female founders in a founding team affects the funding amount. We find that the relative amount of female founders has a negative impact on the funding raised. Furthermore, we observe that founder characteristics have an effect on the funding raised based on the founders' gender. Moreover, we find that gender bias in early-stage funding is less prevalent for serial founders with entrepreneurial experience as female founders benefit three times more than male founders from already having founded a startup. Overall, our study suggests that gender bias exists and is worth to be considered in the context of startup funding. 
\end{abstract}




\end{frontmatter}


\section{Introduction}
\label{sec:Introduction}

Startups play a key role in today's economy. The amount of investments in US-based startups ranging between \$25m and \$100m has consistently increased over the last decade \citep{Rowley.8.3.2019}. 
Determining particularly decisive factors leading to successful startup investments is therefore in great demand. 
Besides startup success prediction, in recent years, the \textit{role of gender} of startup founders has been considered with increasing interest~\citep{Ewens.2020,Lehto.2019}. %
Early studies on gender effects have already revealed differences between genders~\citep{Greene.2001} and confirmed by various studies using data from the U.S. market. 
In this paper, we focus on gender bias with regard to early-stage funding startups. 
The goal of our paper is to answer the question whether gender bias exists among investors by analyzing how the amount of female founders affects the funding amount.

In contrast to related works analyzing gender bias mostly in the US, we focus on the European market and analyze gender bias across European countries. To the best of our knowledge, no other research analyzed gender bias in funding with regard to more than one European country or across all European countries.

Our approach utilizes \textit{Linked Crunchbase},\footnote{http://linked-crunchbase.org} a knowledge graph containing structured information from Crunchbase,\footnote{https://crunchbase.com} an online platform providing information about startups, venture capitalists (VC) and related topics. Since our data set is available in a structured format -- using semantic web standards such as RDF --, our analysis approach based on correlation analysis and regression analysis can utilize the graph-based data model that is enriched with additional information of the Linked Open Data (LOD) cloud, encompassing thousands of freely available knowledge repositories. 
This allows us to consider additional variables (e.g., educational level of founders) in our study besides commonly used variables, such as the startups's industry group or team size. 
Furthermore, our gender analysis shows prevalent effects on startup funding and from which characteristics female founders can benefit (e.g., higher education, serial founder).

Overall, our main contributions are as follows:
\vspace{-0.2cm}
\begin{itemize}
    \item We analyze gender bias concerning European startups using correlation and regression analysis based on a large knowledge graph that is interlinked with the Linked Open Data cloud.
    \item We discuss the results of the gender analysis in detail, revealing prevalent effects on startup funding from female founders' point of view.
\end{itemize}

The rest of our paper is structured as follows: 
We first describe the data set used for our analysis in Section~\ref{sec:dataset}, before outlining our methods for analyzing gender aspects in the startup domain in Section~\ref{sec:methods}. We then present our analysis results in Section~\ref{sec:results} and summarize the main lessons learned in Section~\ref{sec:main-findings}. After describing related work in Section~\ref{sec:related-work}, we conclude the paper in Section~\ref{sec:conclusion}. 
\section{Data Set}
\label{sec:dataset}

\subsection{Linked Crunchbase}
In this paper, we use the knowledge graph \textit{Linked Crunchbase} containing extracted information from the online data platform Crunchbase in a semantically-structured way, providing information about startups, VCs and related topics. Data is entered to Crunchbase via crowdsourcing, i.e. everyone can contribute with information about startups, along with source references (e.g., links to SEC filings or news articles), which are verified by Crunchbase moderators. Crunchbase also provide VCs an incentive to contribute to the database via the \textit{Crunchbase Venture Program}. In exchange for up-to-date data on themselves and their portfolio firms, VCs get certain benefits like discounted access to the database~\citep{Crunchbase.VentureProgram}.

Linked Crunchbase contains information on organizations, investments, and background on leading figures of the worldwide startup economy.
In total, the knowledge graph contains information of about 659,000 organizations, 781,000 people, and 222,000 funding rounds composed of 945,000 investments. In terms of raw size, this data set surpasses many other data sets used previously in related work concerning gender studies. Note, that subsets of Crunchbase have previously been used for gender studies \citep{Raina.2017}.

\begin{figure}
    \centering
    \includegraphics[width=\linewidth]{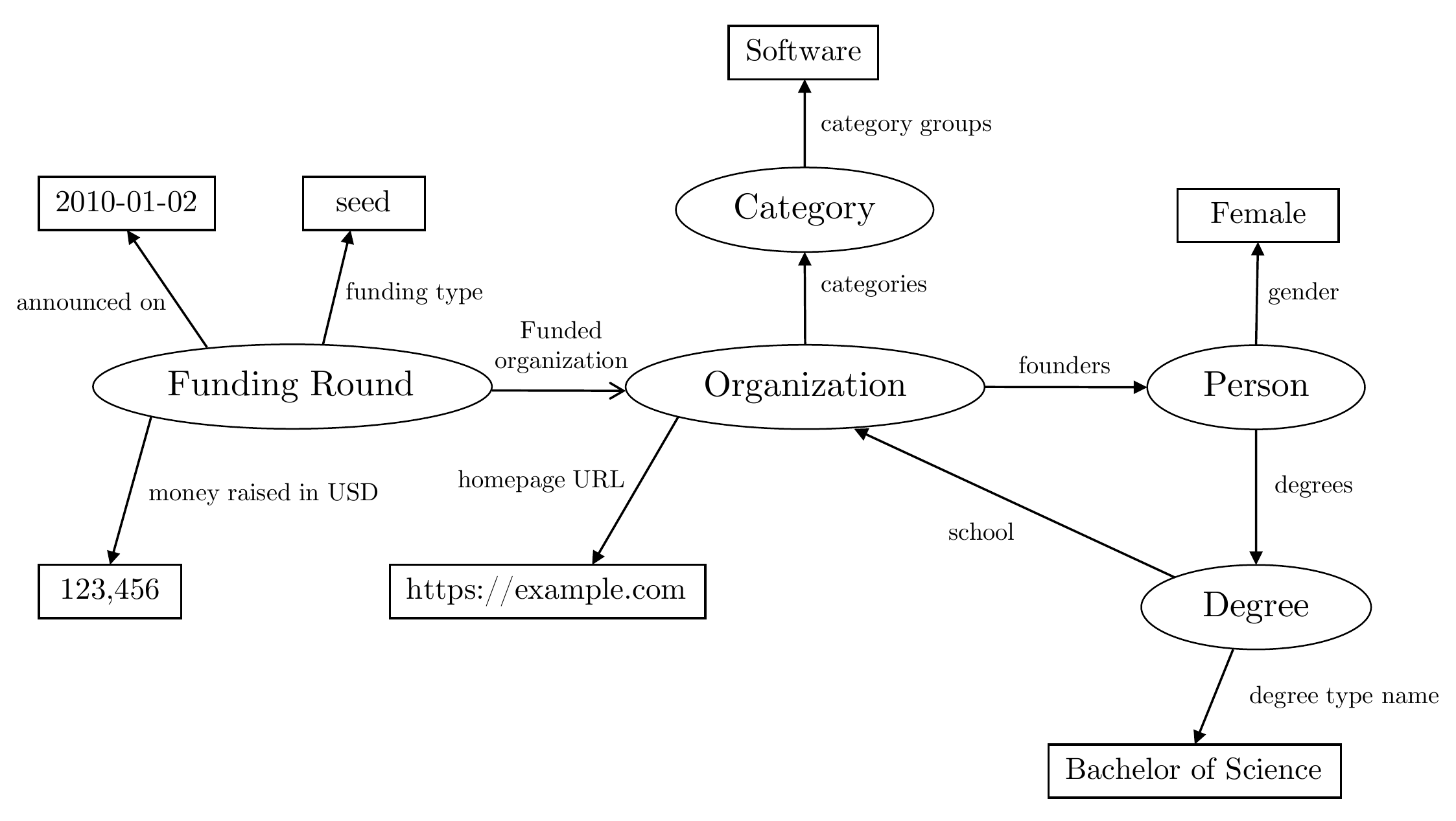}
    \caption{Used Entities and Properties of Linked Crunchbase}
    \label{fig:linked-crunchbase-data}  
\end{figure}

As depicted in Figure \ref{fig:linked-crunchbase-data}, for the analysis, we use all entities of type \textit{funding round}, \textit{organization}, and \textit{person}. A funding round entity provides the funding amount, date of announcement, and a link to the funded organization. It also optionally provides the type of funding, e.g. \textit{seed}, \textit{angel}, or \textit{venture}. An organization provides the founding date, (market) categories, the official homepage URL, and links to its founders. Additionally, it links to its headquarters' address containing its country which we use in our analysis to determine European startups. A person entity yields a person's gender and -- if provided -- links to scholar degrees achieved. These degrees provide information on the degree type (e.g. \textit{B.Sc.}, \textit{PhD}, \textit{MBA}) and a link to the organization entity of the associated school or university. 

\subsubsection{Data Sample}

The data used for our analysis constitutes only a subset of Linked Crunchbase. The analyzed sample of the knowledge graph restricts the data with regard to (1)~the funding date, (2)~the funding amount, and (3)~the location of the headquarter of the individual organization. A majority of organizations on Crunchbase never raised a funding round. Thus, overall, we considered 4,854 
out of the 658,963 organizations.

\textbf{Time period} 
The period examined for our analysis ranges from 2008 to 2018. Even though the data set covers investments from before the '80s until now, the period examined is narrowed down, reducing \textit{survivorship bias}: Since Crunchbase was introduced in July 2007, data from before can be prone to being biased, as it is likely that (early-stage) data of unsuccessful companies from that time is missing. Furthermore, the knowledge represented in Linked Crunchbase was crawled from Crunchbase in mid-2018 and therefore only contains data up until that point in time.

\textbf{Funding amount}
In this analysis, we focus on seed funding rounds for which the amount of money raised by the funded organization is given and surpasses a defined threshold. Some funding rounds in the dataset appear to have incorrect information regarding the amount of dealt financing, which are sometimes unusually low for regular funding rounds, even for early-stage seed funding. This may be due to human errors on the side of Crunchbase or crawl errors while creating the Linked Crunchbase knowledge graph.
Therefore, we only consider startups with at least \$5,000 raised in a funding round. This threshold value is similar to the study of 
\citet{Ewens.2020} based on crowdsourced data of \textit{AngelList},\footnote{https://angel.co} a startup platform for early-stage financing.

\textbf{Geographical Region}
In this paper, we focus on funded companies of European countries allowing 
us to analyze potential imbalances across countries and geographical regions. 
We show in Section \ref{sec:related-work} that most related works analyze US markets.
In comparison, only little research on gender bias has been pursued for the European market, likely due to the availability of US market data sets with a high level of detail.

Crunchbase contains information on worldwide entrepreneurial activity since people and organizations all over the world contribute to the platform. Besides our analysis of the European market and prior US-based startups, it is thus possible to examine also other regions, showcasing the international scope of the data set.

\subsubsection{Data Variables}
\label{sec:variables}

In the following, we present all data variables used in our analysis. While, for example, \textit{team size} or \textit{industry group} are adopted based on properties of the Crunchbase platform, some variables, such as \textit{top university} or \textit{degree}, are derived from properties or result from utilizing another knowledge graph. 

\begin{enumerate}
 \item \textbf{Funding Type}
This optional variable describes the type of funding received by the funded organization, such as \textit{seed}, \textit{venture} or \textit{angel} funding. While for most funding rounds exactly one funding type value is present, some funding rounds also have multiple funding types given, such as \textit{seed} and \textit{angel}. In the latter case, one data point is created for each given funding type. In our analysis, we will only focus on \textit{seed} funding rounds. 

 \item \textbf{Funding Amount}
This variable describes the amount of money that was raised in a funding round. The money raised is the sum of all investments done by all investors involved in this funding round.

 \item \textbf{Industry Group}
\label{sec:industry-group}
In general, the industry group denotes in which industry group an organization operates. Within our analysis, we include the industry group variable to categorize data points for further analysis in order to account for fixed effects that might be present in each industry. For instance, Biotech startups need to raise greater amounts of funding than software startups due to high costs of laboratory equipment.

 \item \textbf{Country}
This variable describes the country of a startup's headquarter, as specified on Crunchbase in August 2018. Hence, we consider only the most recent address and assume that startups did not change the country they are headquartered in.

 \item \textbf{Team Size}
The team size variable stands for the number of founders an organization is associated with. This value is used to normalize team dependent variables. For example, to know the share of female founders in a startup we take the absolute number of female founders in a startup and divide it by the size of the founder team. Additionally, this variable is used to account for random or fixed effects that might depend on the number of founders. 

 \item \textbf{Serial Founder}
The share of serial entrepreneurs in an organization's founding team is indicated by this variable. A serial founder is defined as a person who has already founded at least one other venture previously. The value is defined for each organization-founder pair by determining whether or not there is another organization founded by the same person prior to the examined organization.

 \item \textbf{Degree (B.Sc., M.Sc., PhD, MBA)}
The degree variable displays how many of an organization's founders hold a certain degree. We assume 
that all given degrees are achieved prior to any organization's funding rounds as start and end date of someones degree are often not given.

 \item \textbf{Top University}
This variable describes whether a person attended any university ranked as top 100 worldwide at least once over the years recorded. Times Higher Education (THE) was chosen as the primary data source for university rankings (see Table \ref{tab:top-10-universities}).

 \item \textbf{Female}
This variable depicts how many of an organization's founders are female, as defined by the property \textit{gender}, resulting a share of female founders. Our analysis will primarily revolve around this value.
\end{enumerate}

\begin{table}[tb]
    \centering
    \caption{Top 10 universities 2020 according to the \textit{Times Higher Education} \citep{timeshighereducation.2020}.}
    \label{tab:top-10-universities}
   \resizebox{.95\textwidth}{!}{%
    \begin{tabular}{clrc}
    \hline
        \textbf{Rank} & \textbf{Name (country)} & \textbf{\# Students} & \textbf{Female:Male} \\
        \hline
         1 & University of Oxford (U.K.) & 20,664 & 46 : 54 \\
         2	& California Institute of Technology (U.S.) & 2,240	& 34 : 66 \\
         3 & University of Cambridge (U.K.) & 18,978 & 47 : 53 \\
         4	& Stanford University (U.S.) & 16,135 & 43 : 57 \\
         5	& Massachusetts Institute of Technology (U.S.) & 11,247 & 39 : 61\\
         6	& Princeton University (U.S.) & 7,982 & 45 : 55 \\
         7	& Harvard University (U.S.) & 20,823 &  49 : 51 \\
         8 & Yale University (U.S.) & 12,402 & 50 : 50 \\
         9 & University of Chicago (U.S.) & 13,833 & 46 : 54 \\
         10 & Imperial College London (U.K.) & 16,760 & 38 : 62 \\
     \hline
    \end{tabular}
    }

\end{table}

\subsubsection{Limitations}
\label{analysis:limitations}

In the following, we outline the limitations of the data set and their effects on our analysis:

\begin{itemize}
 \item \textbf{Binary variable for funding}
The question arises why we do not analyze funding bias using a binary variable for funding, i.e. \textit{received funding} -- \textit{received no funding}, instead of the funding amount. Note, that the data set would need to contain both successful funding rounds and rejected funding proposals. However, Crunchbase does not contain information about the latter: We only know who invested in a certain company but not who decided not to invest. Furthermore, it is not known how many attempts a company needed or how often a startup presented its business plan to potential investors until they finally received funding. Overall, given our data set, it is not possible to correlate the probability of receiving any funding at all to the the \textit{founder's gender} or other variables.

 \item \textbf{Target Funding}
Since funding target can be provided on the Crunchbase, an analysis on whether female or male entrepreneurs are more likely to reach their target funding would be possible. 
However, the funding target is mostly missing in the data set. 
Although some companies decide to publish targets for their (prior) funding rounds, there is not enough data available for a statistically significant analysis.
Thus, it is also not possible to correlate the chance of hitting the funding target to founder gender and other variables.

 \item \textbf{Supply-side variables}
Analyzed variables of the analysis can have their origin in supply- or demand-side. In this paper, we focus on the gender on demand-side. 
As we apply a thorough analysis to extract trends as well as influential factors of gender bias itself, an inclusion of possible variables from supply-side goes beyond the scope of this paper. 
\end{itemize}

\section{Methodology}
\label{sec:methods}

To see whether there is a correlation between founder gender and early-stage funding amount, we first apply a \textit{correlation analysis} to our Linked Crunchbase subset. Furthermore, similar to related studies \citep{Ewens.2020}, we also propose a \textit{regression analysis}. This enables us to further understand the exact effect of a founder's gender on the funding amount received.%

\subsubsection{Correlation Analysis}
We use correlation analysis to analyze whether there is a connection between the genders of a company's founders and the amount of funding raised in an early-stage funding round. First, we calculate the correlation between all variables in the sample. Then, we analyze the correlation between the amount of female founders and funding in more detail by examining correlation coefficients across industries and countries, as well as their change over time.

Rank correlation coefficients such as Kendall's rank correlation coefficient \citep{Kendall.1938} (also \textit{Kendall's $\tau$}) or Spearman's correlation coefficient \citep{Spearman.1904} measure the tendency of one variable increasing while the other variable increases as well, without requiring a linear relationship.
We choose \textbf{Kendall's rank correlation coefficient} \citep{Kendall.1938} over Spearman's correlation coefficient since it is less prone to outliers. In addition to this, the coefficient performs better when the sample size is small. 

Formally, the null hypothesis of Kendall's $\tau$ is that there is no trend ($H_0: \tau = 0$). The alternative hypothesis is that there is a trend ($H_1: \tau \neq 0$).
Given that we have the concepts in a temporal order, let $G_i$ be the number of data points after $y_i$ that are greater than $y_i$. Let $L_i$ be the number of data points after $y_i$ that are smaller than $y_i$. Then, the Kendall's $\tau$ coefficient is calculated as
\begin{equation}
 \tau = 2S/n(n-1)
\end{equation}
and $S$ is thereby defined as
\begin{equation}
 S = \sum_{i=1}^{n-1}{(G_i - L_i)}
\end{equation}
and corresponds to the the sum of the differences between $G_i$ and $L_i$ along the time series.
Since we are dealing with a sufficiently large number of time slots $n$, we can assume normal distribution for the test statistic $z$ 
and write 
\begin{equation}
 z = \frac{\tau}{\sqrt{2(2n+5)/9n(n-1)}}
\end{equation}

\subsubsection{Regression Analysis}

To further understand the interaction between founder gender and funding received, we use regression analysis. 
As applied in related works on gender effects in startup investments (e.g., \citet{Ewens.2020}, \citet{Lins.2016}), we use \textbf{ordinary least squares (OLS) regression} for estimating the linear relationship between funding size and founder-level variables. 
\begin{equation}
    raised_i = \alpha + \beta Female_i + \delta X_i + \epsilon_i
\end{equation}
where $i$ indexes the funding round examined, $Female$ represents the relative share of female founders of the funded organization and $X$ is a vector of additional founder-level variables. The dependent variable $raised$ represents the money raised in a funding round (in US Dollar).
In order to account for fixed effects, which may, for instance, occur across industries or countries, the regression is applied on several groups in isolation. Possible fixed effects to include are \textit{year}, \textit{team size}, \textit{industry} and \textit{country}.

Similar to other studies, we analyze the sample period as a whole. This reveals whether gender affects the amount of received funding and to what extent. 
Additionally, we apply the regression separately over time to examine how the influence of the team variables ($\beta$) changes over the years. To the best of our knowledge, there is no study analyzing the gender effect over time. Furthermore, we use the regression analysis for comparing male and female founders in order to examine differences in how the founder characteristics affect the funding raised differently across genders. We do this by analyzing startups with only one founder and then apply the OLS regression for both male and female startups separately.

\section{Analysis Results}
\label{sec:results}

In the following, we outline both descriptive statistics and baseline findings of our analysis as well as describe the results of the correlation and regression analysis in detail.

\subsection{Basic Findings}
\label{sec:basicfindings}

The gender composition of the founders in the analyzed subset of Linked Crunchbase is presented in Table \ref{tab:results-baseline-gender-distribution} and compared to other data sets. Similar to prior studies \citep{Ewens.2020,Brush.2018,Lins.2016}, we observe that startups are  less frequently founded by women.
Furthermore, the gender distribution in our European subset is similar to US-based data sets. We find a significant difference between Crunchbase, a data set mostly containing already funded organizations, and AngelList, a platform for startups actively seeking early-stage financing. In our sample, the share of startups with at least one female founder is 15.7\%, as compared to AngelList with a share of 20.9\%. Hence, we conclude that female founders are less successful in raising early-stage funding than male founders.

\begin{table}[tb]
    \centering
    \caption{Gender distribution of startups across different data sets.} 
    \label{tab:results-baseline-gender-distribution}
    \resizebox{\textwidth}{!}{%
    \begin{tabular}{lccccc}
        \hline
         & \textbf{Crunchbase} & \textbf{AngelList} & \textbf{VentureSource} & \textbf{Pitchbook} & \textbf{KfW/ZEW} \\
        \hline
        Male-only      & 85.3\% & 79.1\% & 87.6\% & 86\% & 82.5\% \\
        Female-involved & 15.7\% & 20.9\% & 17.3\% & 14\% & 17.5\% \\
        \hline
    \end{tabular}
    }
    \end{table}
    
\begin{table}[tb]
\centering
 \caption{Descriptive summary statistics for male-only and female-involved startups.} 
 \label{tab:results-baseline-variable-distribution}
 \resizebox{\textwidth}{!}{%
 \begin{tabular}{lrrrrrr} 
 \hline
              & \multicolumn{3}{c}{\textbf{male-only startups}} & \multicolumn{3}{c}{\textbf{startups with female founders}}  \\
              & \textbf{mean}    & \textbf{median}  & \textbf{std. dev.}          & \textbf{mean}    & \textbf{median}  & \textbf{std. dev.}                       \\ 
\hline
Team size     & 1.989   & 2       & 1.039              & 2.457   & 2       & 1.240                           \\
Serial        & 0.222   & 0       & 0.556              & 0.215   & 0       & 0.572                           \\
B.Sc.      & 0.224   & 0       & 0.528              & 0.395   & 0       & 0.660                           \\
M.Sc.        & 0.294   & 0       & 0.582              & 0.531   & 0       & 0.748                           \\
MBA           & 0.070   & 0       & 0.277              & 0.097   & 0       & 0.343                           \\
PhD        & 0.055   & 0       & 0.257              & 0.089   & 0       & 0.393                           \\
Top university     & 0.213   & 0       & 0.502              & 0.353   & 0       & 0.641                           \\
Funding (USD) & 655,069 & 265,000 & 1,228,870          & 532,926 & 162,429 & 921,093                         \\
\hline
\end{tabular}
}
\end{table}

\begin{figure}[tb]
    \centering
    \includegraphics[width=0.7\linewidth]{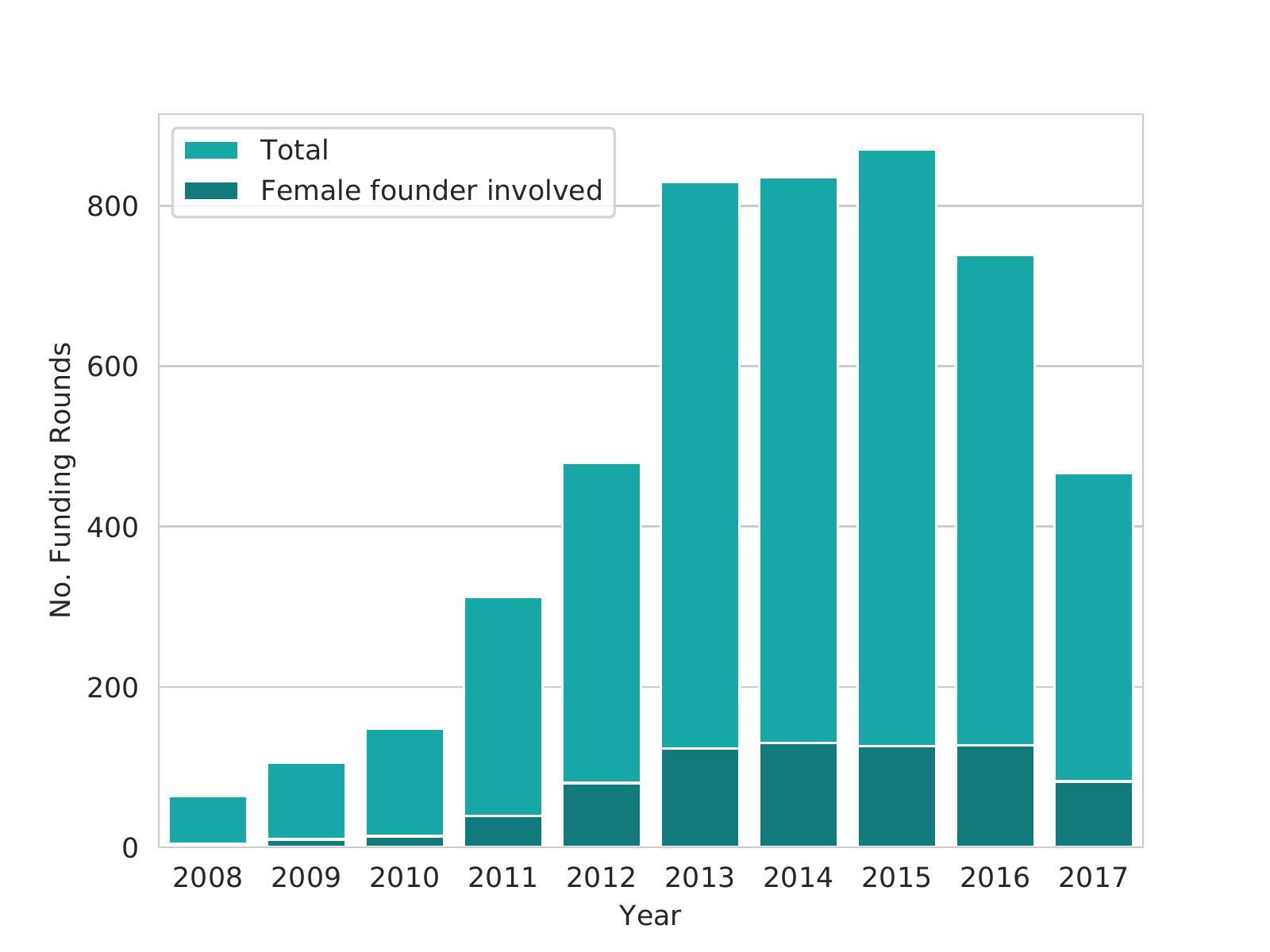} %
    \caption{Distribution of funding rounds over time. }%
    \label{fig:results-baseline-years}
\end{figure}

\textbf{Summary statistics} In Table \ref{tab:results-baseline-variable-distribution} the descriptive summary statistics are presented, displaying the distribution of the founder-level variables in our subset, including funding amount raised (in USD). Here the characteristics for startups with male-only founding teams are compared to startups with female founder involvement. We observe that, on average, (i)~\textit{female-involved startups raise smaller early-stage rounds} than male-only startups do, and 
(ii)~\textit{female-involved startups are more likely to have higher education}. Hence, the share of founders with an academic degree are greater at the female-involved startup. Moreover, the share of top university graduates is also greater.

\begin{figure}[tbp]
    \centering
    \includegraphics[width=0.82\linewidth]{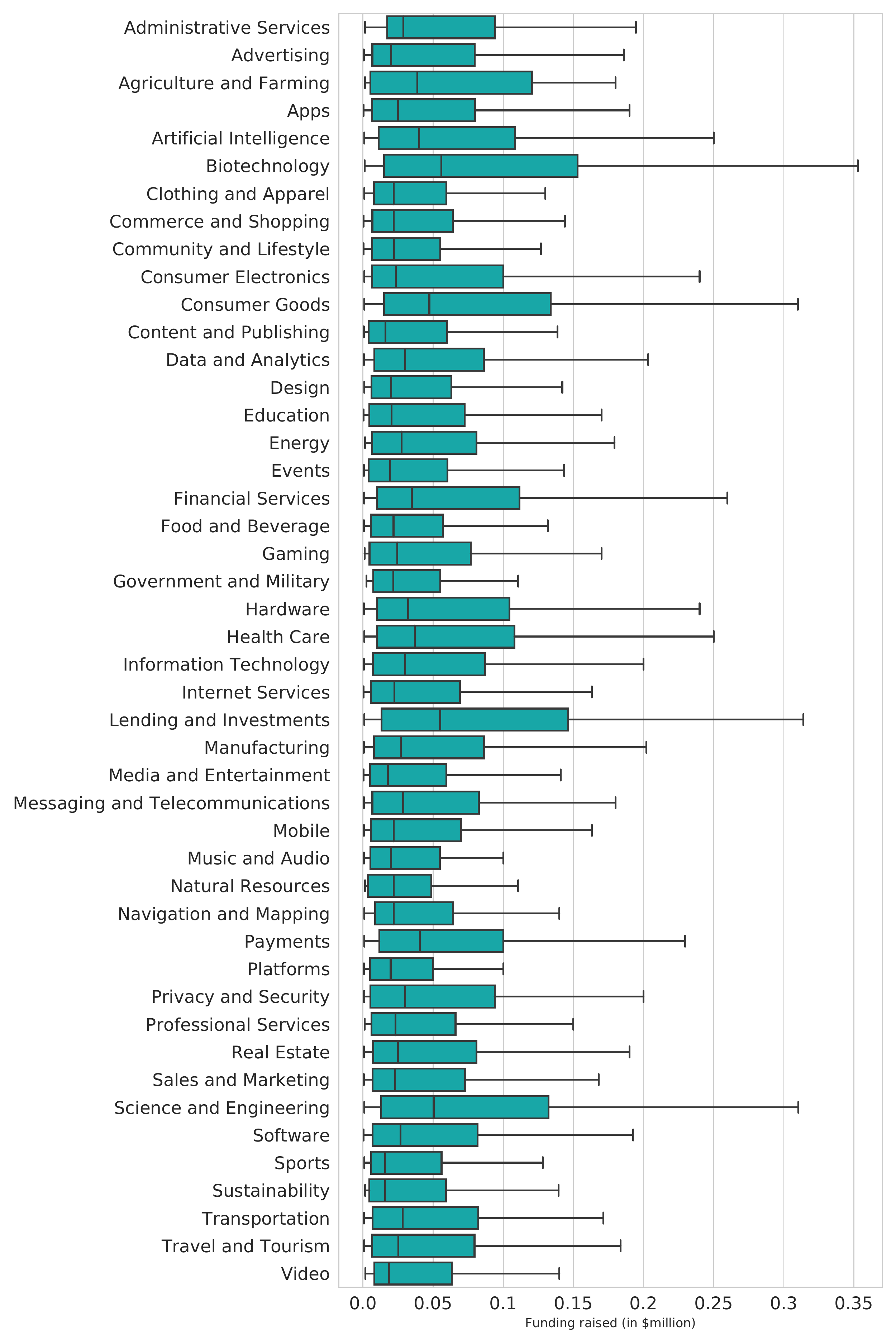}
    \caption{Distribution of average funding across industries.}
        \label{fig:results-baseline-industry-boxplot}
\end{figure}

At first sight this observation seems contradictory as startups with better educated founders (i.e. more intellectual capital) should receive higher valuation.
In Section \ref{results:regression}, we apply a regression analysis to further elaborate on this effect and conclude that work experience might be outweighing higher education.

\begin{figure}[tbp]
    \centering
    \includegraphics[width=0.8\linewidth]{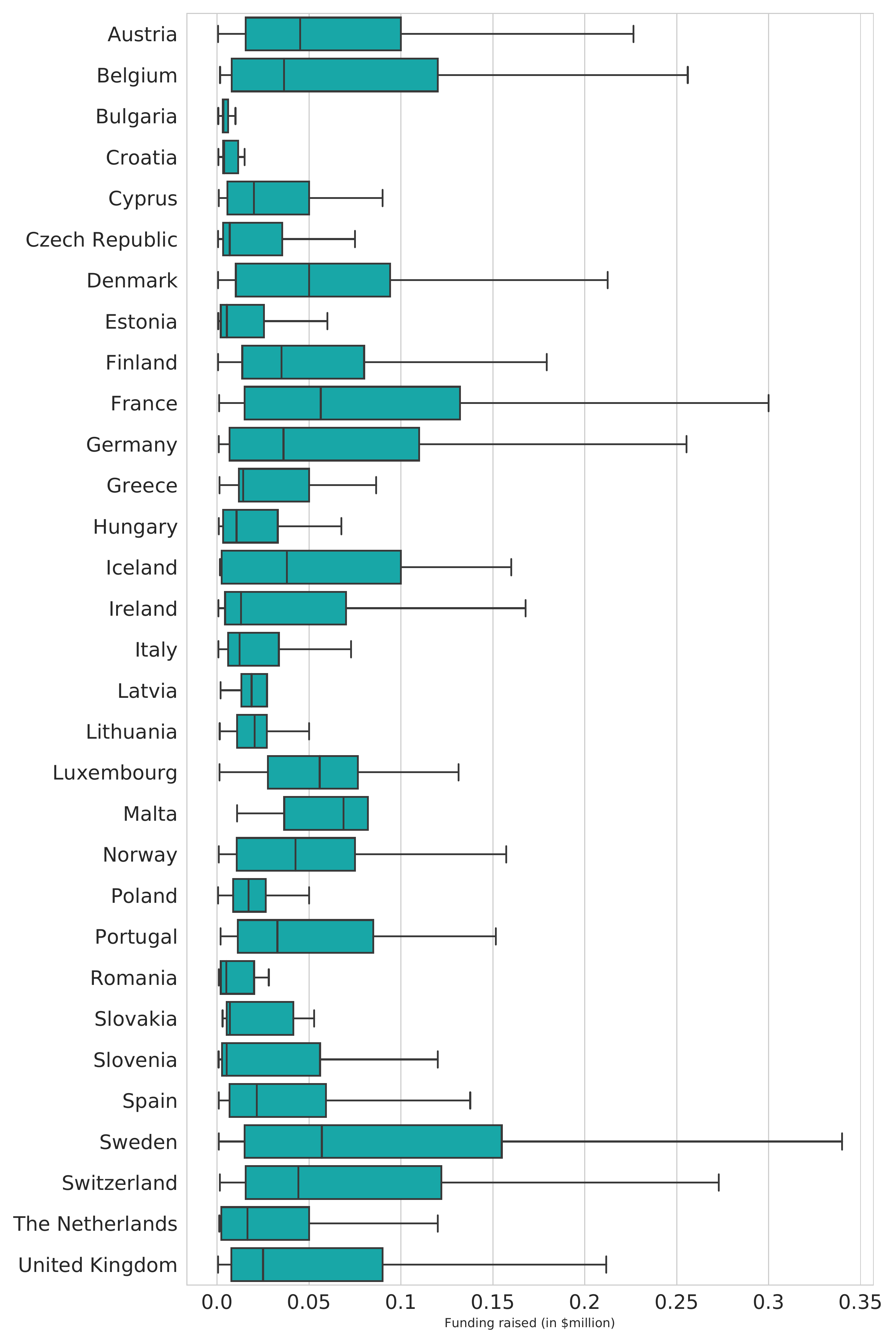}
    \caption{Distribution of average funding across countries.}
    \label{fig:results-baseline-country-boxplot}
\end{figure}

\textbf{Funding rounds over time} In Figure \ref{fig:results-baseline-years}, the number of funding rounds announced are presented, per year for the period between 2008-2017. We see that the funding rounds are not evenly distributed, with highest observations in the years 2012-2017. We assume that this due to the fact that Crunchbase was first released in 2008 and the years after not well-known yet, in particular outside of the US.

\textbf{Funding amounts across industries} Figure \ref{fig:results-baseline-industry-boxplot} displays a boxplot of early-stage funding amounts raised across all industry groups examined. There are clear disparities between different industries in the average funding amount. We observe that startups in the biotech and healthcare raise the most money, while the sports and app development industry raise the least.
This underlines the necessity to consider fixed effects in our analysis.

\textbf{Funding amounts across countries} In Figure \ref{fig:results-baseline-country-boxplot} the funding amount across European countries is depicted. We observe even greater differences in early-stage financing raised. Switzerland is by far the country with the highest raised funding amount per startup. In contrast, startups in countries like Poland, Spain, and Greece, raise significantly less money. This suggests a connection between the \textit{GDP per capita} of a country and the funding raised by a startup headquartered in the respective country. Thus, fixed effects across countries must also be considered.

\subsubsection{Trends over the Years}
\begin{figure}[tb]
    \centering
    \includegraphics[width=0.67\linewidth]{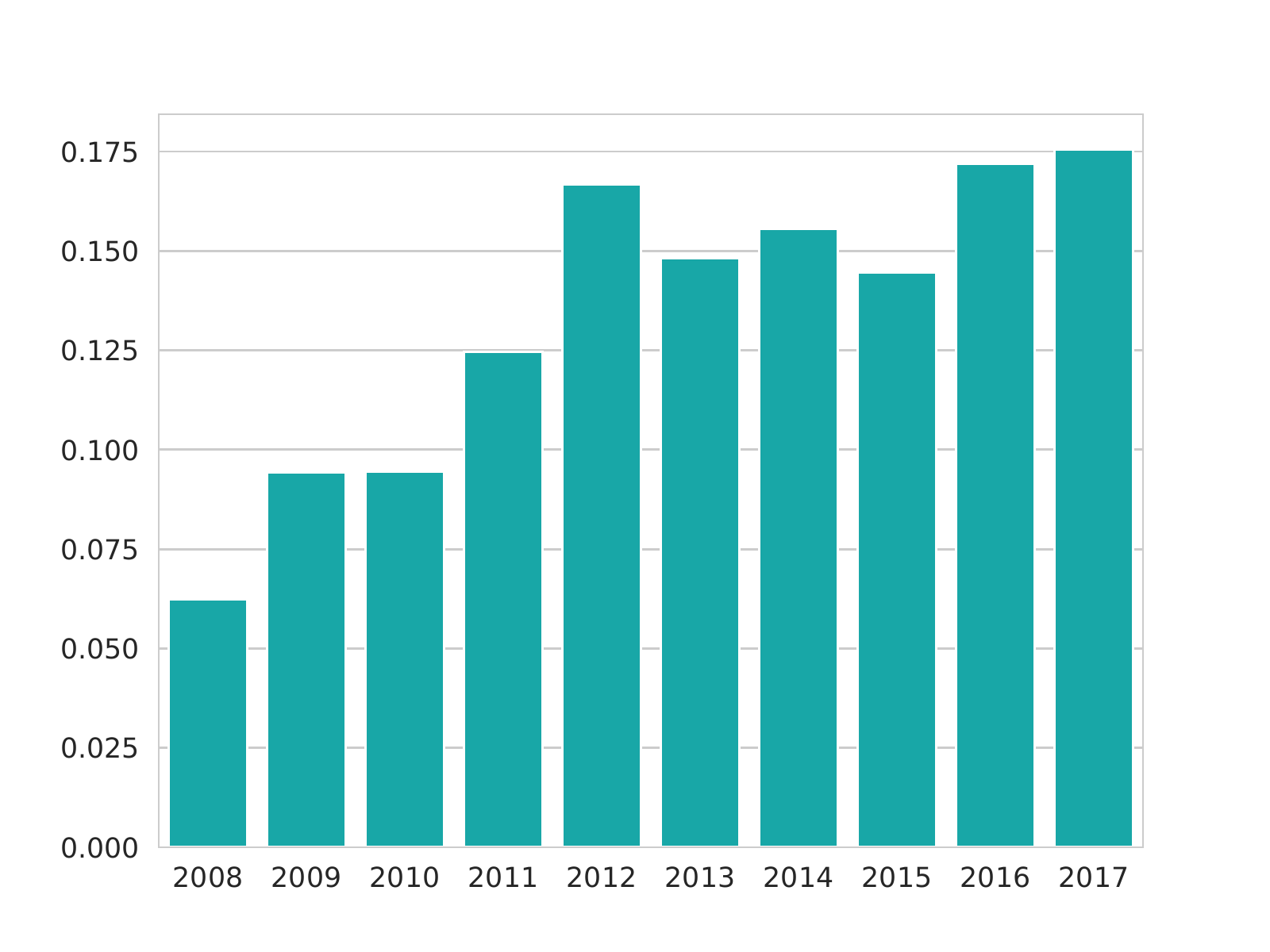}
    \caption{Share of funding rounds with female-involved startups in Europe.}
    \label{fig:results-baseline-year-female}
\end{figure}

In Figure \ref{fig:results-baseline-year-female} the share of funded startups with at least one female founder is plotted against time, a clear trend is visible: The ratio of female-involved startups per year increased from 6.62\% in 2008 up to 17\% in 2017. The results also line up with the findings from 
\citet{Brush.2018} reporting an average participation of female-led businesses at 15\% in the period of 2013-2015. Although the ratio between startups with female founders and male-only startups is still far from balanced, the continuous positive trend of funded startups let anticipate a gender-equal future in entrepreneurship.

\begin{figure}[tb]
    \centering
    \includegraphics[width=0.72\linewidth]{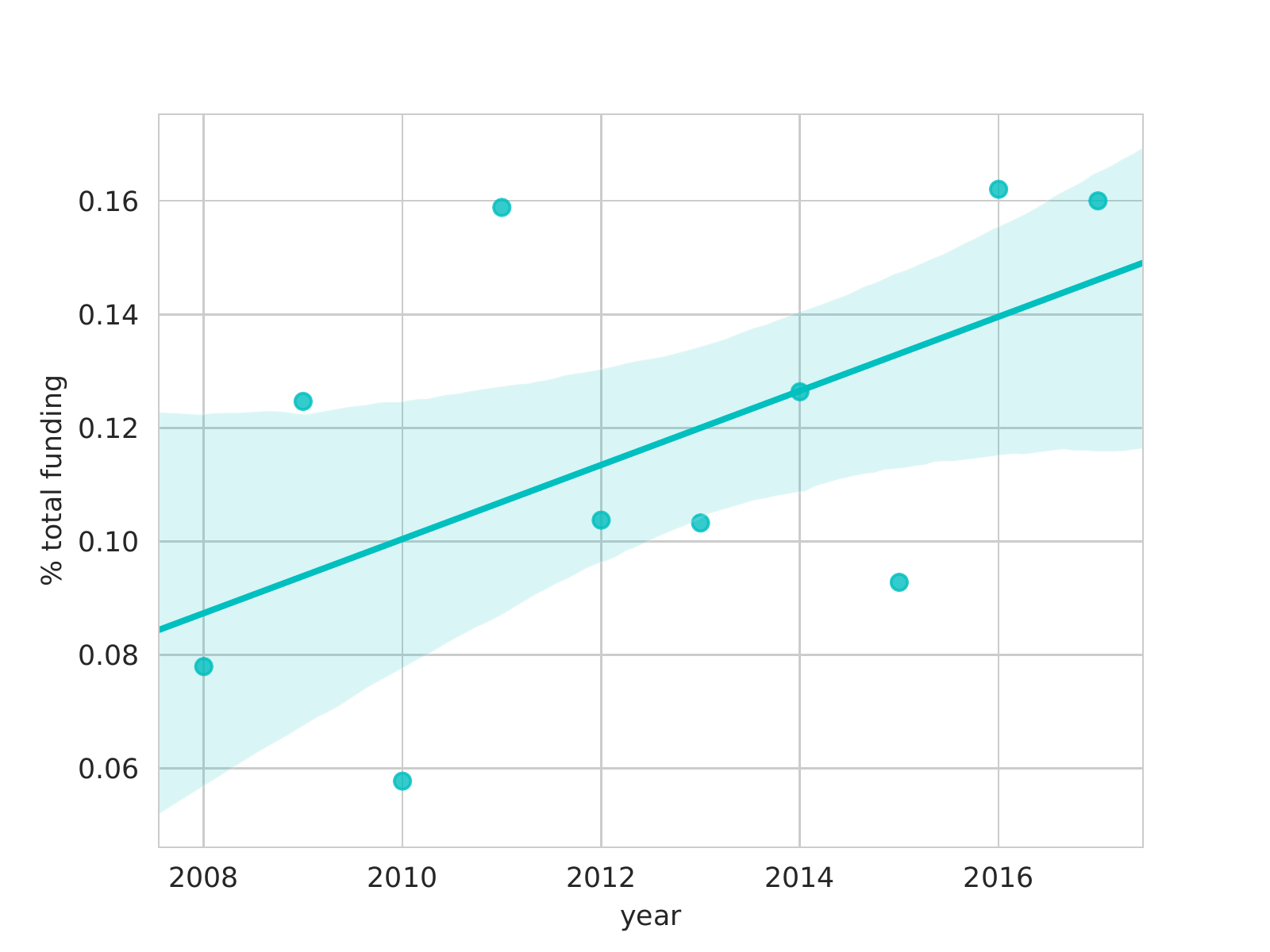}
    \caption{Distribution of USD invested in female-involved organizations relative to total USD invested per year, including an OLS regression with a 90\% confidence interval.}
    \label{fig:results-baseline-year-female-dollar}
\end{figure}

Furthermore, similar findings can be derived from the share of total amount funded per year, as depicted in Figure \ref{fig:results-baseline-year-female-dollar} with an additional linear regression and a 90\% confidence interval. Although the values are volatile, an overall positive trend can be observed in the share of dollars invested in female-involved startups over the examined period. The total funding received started at 6\% and increased to 16\% in 2017.

\subsubsection{Trends in Industry Groups and Countries}
\begin{figure}
    \centering
    \includegraphics[width=0.8\textwidth]{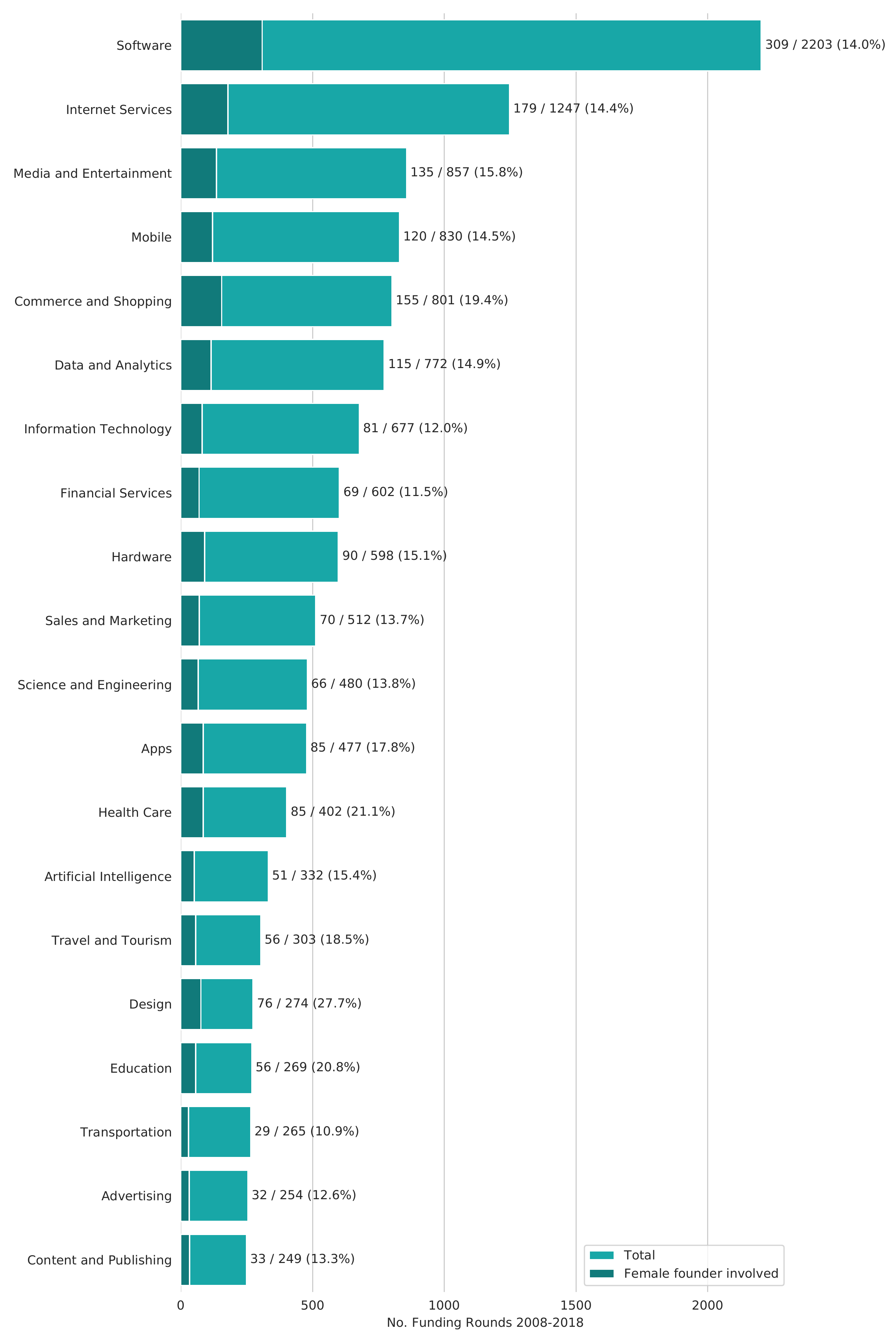}
    \caption{Distribution of examined funding rounds across industries (top 20)}
    \label{fig:results-baseline-industry-distribution}
\end{figure}

\begin{figure}
    \centering
    \includegraphics[width=0.82\textwidth]{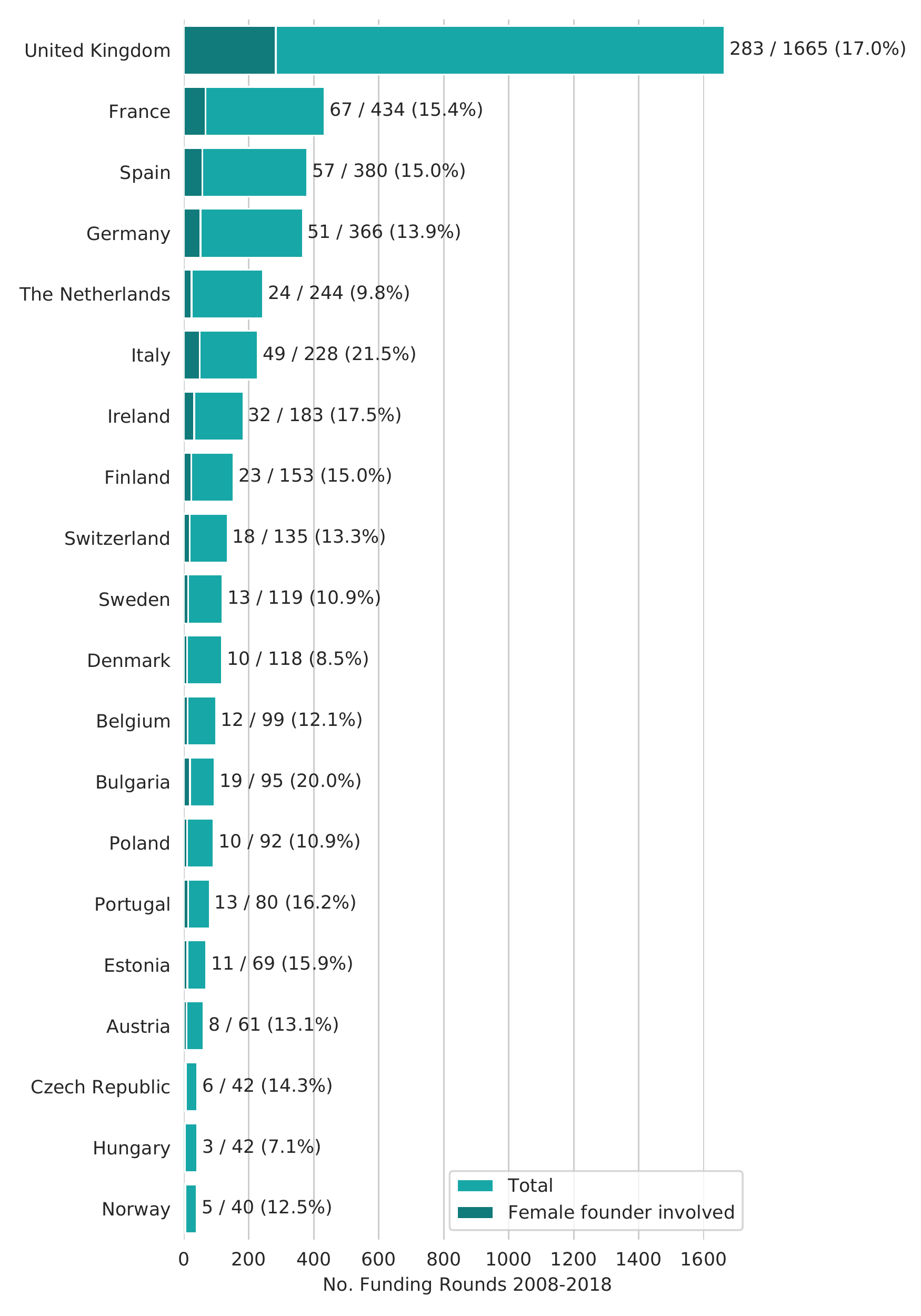}
    \caption{Distribution of examined funding rounds across countries (top 20)}
    \label{fig:results-baseline-country-distribution}
\end{figure}

Considering industry groups (Figure \ref{fig:results-baseline-industry-distribution}) and countries (Figure \ref{fig:results-baseline-country-distribution}) it can be observed that the data is not evenly distributed across these variables.

\textbf{Industry Groups} Examining the distribution of funding rounds across different industry groups (see Figure \ref{fig:results-baseline-industry-distribution}), we find significant differences regarding female-involved startups. The three industries with the highest share of founding rounds with female founders are \textit{Clothing and Apparel} (41.5\%), \textit{Design} (27.7\%) and \textit{Consumer Goods} (23.8\%). These results reflect the claims by 
\citet{Brush.2018} that the gender gap varies between industries as women-led ventures are more prominent in the service sector.

\textbf{Countries} Figure \ref{fig:results-baseline-country-distribution} displays the variance in the ratio of female-involved with respect to the country they are headquartered in. Italy and Bulgaria have a share of 21.5\% and 20\% respectively compared to Denmark's 8.5\% 
of female-involved startups in the data set.

\subsection{Correlation Analysis}

\subsubsection{Overall Correlations}

\begin{figure}[tbp]
    \centering
      \includegraphics[width=0.55\linewidth]{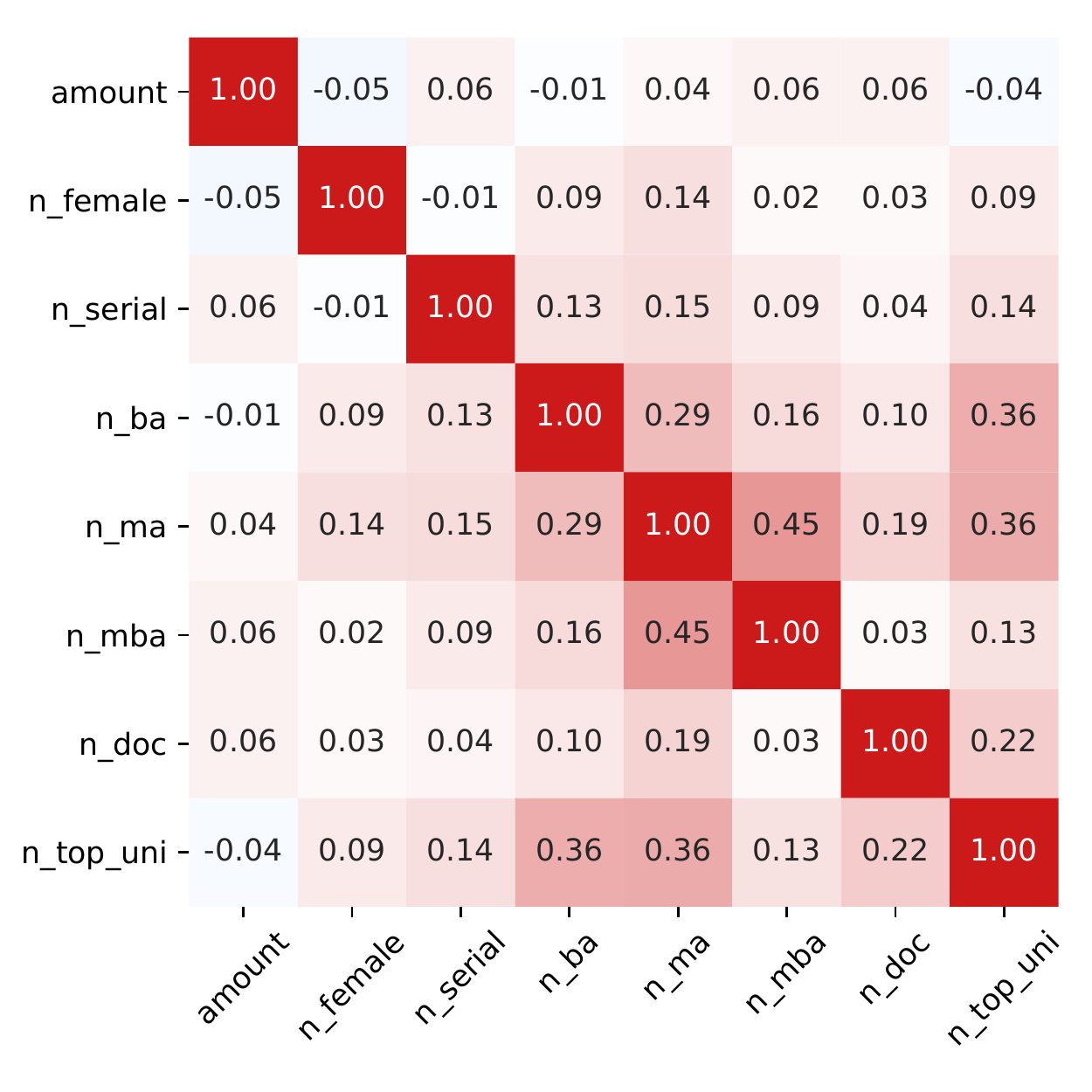}
      \includegraphics[width=0.55\linewidth]{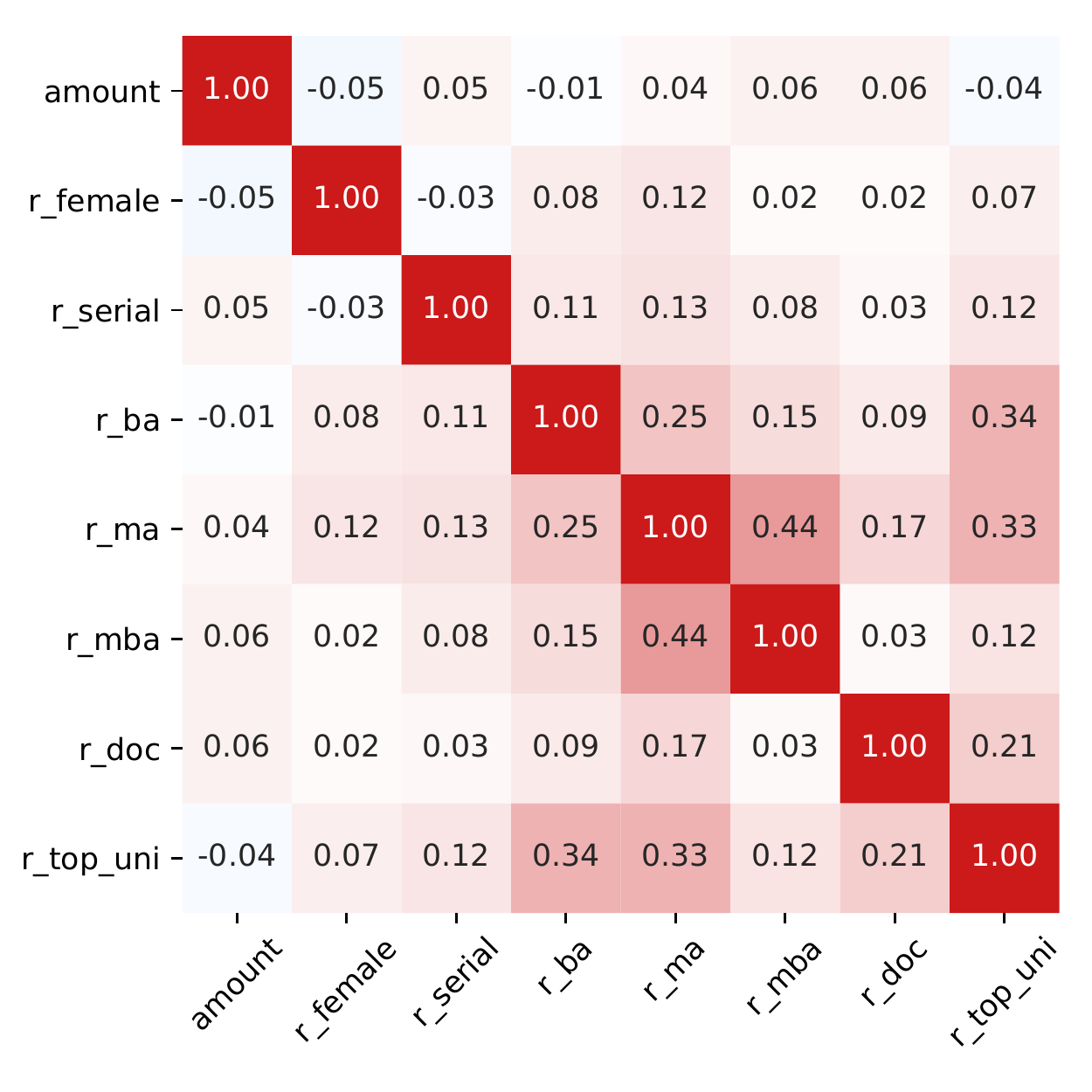}
    \caption{Kendall correlation coefficients between each variable, calculated across the whole sample. (a) uses variables displaying the total number of affected team members (b) uses variables relative to team size}
    \label{fig:results-corr-all-correlations}
\end{figure}

When examining the correlation between all variables across the whole subset, we find mostly weak correlation (see Figure~\ref{fig:results-corr-all-correlations}). Apart from that, there is no correlation between the funding amount and the examined team variables. The two smallest correlation coefficients regarding funding amount are found in the amount of female founders (-0.05) and top university graduates (-0.04). The same applies to their relative shares. As discussed in Section \ref{sec:basicfindings}, the slightly negative correlation coefficient shows that graduating from a elite university exhibits not necessarily a positive signal to investors. To this end, we assume that fixed effects regarding industries and years might affect the funding amount to a higher degree than the founders' gender. Hence, in the following sections, we will examine this relationship in more detail by looking at industries separately. Furthermore, the following analysis will proceed with relative team variables only, as the difference between using relative and absolute variables can be neglected in the analyzed subset.

\subsubsection{Correlations Considering Fixed Effects}

In the following, we analyze the correlation between female founder rate and funding amount and whether it changes over time.
First, we will investigate this relationship across industry groups, then continue with investigating differences between countries. Note that we will only include the top 10 industries and countries respectively, based on the number of associated funding rounds.\footnote{Splitting the data set based on both industry and time greatly reduces the size of the observed groups and statistical significance cannot be ensured for smaller groups}
Overall, a trend of the correlation coefficients cannot be observed as values change inconsistently between -0.04 and -0.08 (see Tables \ref{tab:results-corr-industry} and \ref{tab:results-corr-country}).

\definecolor{red4}{rgb}{0.545,0.000,0.000}
\definecolor{green4}{rgb}{0,0.392,0}
\definecolor{blue4}{rgb}{0.502,0,0.502}
\newcommand{\mcdr}{\color{red4}}
\newcommand{\mcdg}{\color{green4}}
\newcommand{\mcdb}{\color{blue4}}
\newcommand{\mcdblack}{\color{black}}
\newcommand{\mcdrx}{\textcolor{red4}}
\newcommand{\mcdgx}{\textcolor{green4}}
\newcommand{\mcdbx}{\textcolor{blue4}}

\newcommand\SignificanceColored 
 {\mcdgx{* $p < 0.10$}, \mcdbx{ ** $p < 0.05$}, \mcdrx{*** $p < 0.01$}}

\begin{sidewaystable}   
\centering
 \caption{Female-Funding correlation across industries. Kendall-Tau correlation coefficients between funding amount and proportion of female founders. Significance: \SignificanceColored.}%
 \label{tab:results-corr-industry}
\resizebox{\textwidth}{!}{%
\begin{tabular}{lSSSSSSSSSS|S}
\textbf{Industry}       & \textbf{2008}  & \textbf{2009}  & \textbf{2010}  & \textbf{2011}  & \textbf{2012}  & \textbf{2013}  & \textbf{2014}  & \textbf{2015}  & \textbf{2016}  & \textbf{2017}  & \textbf{Total}   \\ 
\hline
Software                & 0.15           & 0.05           & 0.0            & 0.04           & \mcdg -0.09*      & \mcdb -0.09**     & \mcdb -0.1**      & \mcdb -0.09**     & -0.02          & -0.03          & \mcdr -0.06***      \\
Internet Services       & 0.25           & -0.03          & -0.11          & 0.01           & 0.02           & \mcdb -0.12**     & -0.04          & -0.05          & -0.03          & \mcdb -0.19**     & \mcdb -0.05**       \\
Media and Entertainment & 0.02           & -0.19          & 0.07           & -0.1           & -0.07          & -0.07          & 0.05           & -0.07          & 0.08           & 0.08           & -0.03            \\
Mobile                  & —              & -0.25          & -0.2           & 0.06           & -0.09          & 0.05           & \mcdb -0.14**     & -0.05          & -0.09          & 0.1            & \mcdg -0.05*        \\
Commerce and Shopping   & 0.09           & -0.12          & 0.0            & 0.05           & -0.03          & -0.02          & -0.06          & -0.09          & 0.04           & -0.04          & -0.03            \\
Data and Analytics      & —              & -0.11          & —              & 0.05           & -0.07          & -0.06          & -0.1           & -0.06          & 0.05           & \mcdg -0.15*      & -0.04            \\
Information Technology  & —              & 0.04           & —              & 0.1            & -0.07          & -0.12          & -0.12          & -0.01          & 0.06           & \mcdg -0.15*      & \mcdg -0.06*        \\
Financial Services      & —              & —              & -0.02          & 0.14           & 0.01           & -0.03          & -0.11          & 0.01           & \mcdb -0.21**     & 0.0            & -0.02            \\
Hardware                & —              & —              & -0.31          & 0.12           & \mcdb -0.22**     & \mcdb -0.24**     & \mcdg -0.14*      & \mcdb -0.18**     & -0.03          & -0.09          & \mcdr -0.14***      \\
Sales and Marketing     & —              & 0.02           & 0.25           & -0.18          & -0.01          & 0.1            & -0.07          & 0.06           & -0.15          & -0.02          & -0.02            \\ 
\hline
\textbf{Total}          & 0.01           & -0.06          & -0.04          & 0.01           & \mcdb -0.04**     & \mcdr -0.06***    & \mcdr -0.08***    & \mcdr 0.07***     & \mcdg -0.03*      & \mcdb -0.04**     & \mcdr -0.05***     
\end{tabular}
}

\vspace{1cm}

\caption{Female-Funding correlation across countries. Kendall-Tau correlation coefficients between funding amount and proportion of female founders. Significance: \SignificanceColored.}%
\label{tab:results-corr-country}
\resizebox{\textwidth}{!}{%
\begin{tabular}{lSSSSSSSSSS|S} 
\hline
\textbf{Country} & \textbf{2008}                           & \textbf{2009}                      & \textbf{2010}                           & \textbf{2011}             & \textbf{2012}                           & \textbf{2013}                           & \textbf{2014}                      & \textbf{2015}                           & \textbf{2016}                           & \textbf{2017}                           & \textbf{Total}             \\ 
\hline
United Kingdom   & -0.04                                   & -0.08                              & \mcdb -0.14** & 0.02                      & \mcdb -0.08** & \mcdr -0.1***                & \mcdr -0.09***          & \mcdr -0.09***               & -0.03                                   & \mcdg -0.06*      & \mcdr -0.06***  \\
France           & \textemdash                                     & -0.05                              & \mcdg -0.16*      & \mcdr -0.21*** & \mcdr -0.26***               & -0.03                                   & -0.0                               & 0.01                                    & 0.07                                    & \mcdr -0.29***               & \mcdr -0.07***  \\
Spain            & \textemdash                                     & -0.36                              & -0.08                                   & -0.04                     & \mcdr -0.16***               & -0.03                                   & -0.06                              & \mcdr 0.25***                & -0.04                                   & -0.0                                    & -0.02                      \\
Germany          & \textemdash                                     & \mcdg -0.21* & \mcdg 0.23*       & \mcdr 0.44***  & \mcdr -0.34***               & -0.02                                   & \mcdg -0.09* & 0.07                                    & \mcdr 0.18***                & 0.11                                    & 0.01                       \\
The Netherlands  & \mcdb 0.47**  & \textemdash                                & \textemdash                                     & \textemdash                       & 0.03                                    & \mcdr -0.19***               & 0.1                                & \mcdr -0.27***               & -0.07                                   & \mcdb -0.22** & \mcdr -0.13***  \\
Italy            & \textemdash                                     & \textemdash                                & \textemdash                                     & \mcdr 0.42***  & \mcdb 0.21**  & \mcdb -0.16** & \mcdr -0.2***           & -0.01                                   & -0.09                                   & \mcdr 0.31***                & -0.01                      \\
Ireland          & \mcdb -0.51** & \textemdash                                & -0.3                                    & \mcdr -0.34*** & 0.11                                    & \mcdb -0.18** & 0.09                               & \mcdb -0.24** & -0.08                                   & 0.1                                     & -0.05                      \\
Finland          & \textemdash                                     & \textemdash                                & \textemdash                                     & -0.17                     & \textemdash                                     & \mcdr -0.45***               & \mcdr -0.24***          & \mcdr -0.34***               & -0.05                                   & -0.18                                   & -0.02                      \\
Switzerland      & \textemdash                                     & \textemdash                                & \textemdash                                     & -0.2                      & \mcdr -0.67***               & \mcdr -0.37***               & \mcdr -0.45***          & -0.1                                    & \mcdb -0.22** & \mcdr -0.41***               & \mcdr -0.22***  \\
Sweden           & \textemdash                                     & \textemdash                                & \textemdash                                     & \mcdr 0.6***   & 0.14                                    & \textemdash                                     & \mcdg 0.2*   & 0.09                                    & 0.0                                     & -0.08                                   & 0.03                       \\ 
\hline
Total            & 0.01                                    & -0.06                              & -0.04                                   & 0.01                      & \mcdb -0.04** & \mcdr -0.06***               & \mcdr -0.08***          & \mcdr 0.07***                & \mcdg -0.03*      & \mcdb -0.04** & \mcdr -0.05*** 
\end{tabular}
}
\end{sidewaystable}

\textbf{Industry Groups}
Looking at the correlations calculated across industries (see Table \ref{tab:results-corr-industry}), we find that coefficients do vary between industries. 
Unfortunately, 
many values in the table are not statistically significant due to the small observation sizes. 
Considering only statistical significant values, correlation coefficients range between a minimum of -0.14 for \textit{hardware} and a maximum of -0.05 for \textit{mobile}.
Including only statistically significant coefficients, the strongest correlation over the years can be found in the \textit{hardware} industry: In 2012 the coefficient peaks at -0.24.
Moreover, it is interesting that -- although small -- all statistically significant (i.e., with a p-value of less than 0.1) correlation coefficients are negative. This in itself suggests that some kind of inequality must be present.

Furthermore, examining the change of the coefficients over time, we can see a trend in the \textit{hardware} industry towards independence between the ratio of female founders and the funding amount. This signals that women entrepreneurs seem to move more and more into a position equal to men. Especially in the case of the \textit{hardware} industry, it suggests that this kind of biased business gets more and more women-friendly.

 \textbf{Countries} Considerable differences exist between countries, as shown in Table \ref{tab:results-corr-country}. Calculating the correlation coefficients for each country yields promising and statistically significant results compared to industry groups. 
Note that many values between 2008 and 2010 are missing due to an insufficient sample size since Crunchbase was launched in 2007 and not yet prominent outside the US. Therefore, data on many EU startups are rather scarce for the years shortly after.

The lowest correlations coefficients (i.e., indicating a low negative correlation) for all years are present for Switzerland (-0.22) and the Netherlands (-0.13). Calculations for other countries such as Germany and Italy result in negligible correlations with coefficients near zero. Highest correlation coefficients are also found near zero. Since the significance of these small values is not given as all p-values are above 0.1, no clear finding can be derived from these specific values.

Furthermore, examining the countries over the course of time, we see even more interesting results. Switzerland consistently exhibits noticeable negative correlations over the years, with a peak correlation coefficient of -0.67 in 2012. Interestingly, all years with a statistically significant correlation coefficients show a low to moderate negative correlation between the amount of female founders and funding received. Similarly, Finland exhibits low to moderate negative correlations, especially in the years from 2013 to 2015. This finding aligns with the research of 
\citet{Lehto.2019} stating that Finnish VCs are even more male-dominant than in the United States. This is unexpected for a country promoting itself as a \textit{gender equality pioneer} \citep{Finland.Gender-Equality-Pioneer}.
In contrast, Sweden shows positive correlations between the funding amount and share of female founders in 2011 and 2014. The greatest correlation measured across all countries is 0.60, exhibited in 2011 by Sweden.
Moreover, Germany's correlation coefficients seem very volatile, with extremes on both sides with values ranging from 0.44 in 2011 to -0.34 the following year.

\subsection{Regression Analysis}
\label{results:regression}

In the following, we apply an OLS regression to analyze the connection between gender and funding amount and to determine how a specific variable affects the funding amount between two similar startups.
First, we examine how general founder characteristics like \textit{gender} and \textit{educational background}  affect the funding amount, followed by an analysis of startups founded by a single person as these contribute to the majority of the sample. In this way, we are able to determine how \textit{gender} influences other present biases.

\subsubsection{The Impact of Founder Characteristics on Funding Received}
All results discussed in the following are shown in Table \ref{tab:results-reg-all}. As founder characteristics we evaluate the impact of \textit{gender}, \textit{educational background} and experience as being \textit{serial founders} on received funding amount.

\begin{table}[tb]
\centering
    \caption{Funding amounts depending on team variables. 
        Standard errors are reported in parentheses. Significance: \SignificanceColored.}
    \label{tab:results-reg-all}
    \begin{footnotesize}
\begin{tabular}{lccc} 
\hline
\textbf{Variable}                & \textbf{(1)}                                                  & \textbf{(2)}                                                  & \textbf{(3)}                                                   \\ 
\hline
\textbf{Intercept}      & \begin{tabular}[c]{@{}c@{}}\mcdr 703,300***\\(260,000)\end{tabular} & \begin{tabular}[c]{@{}c@{}}\mcdb 556,200**\\(258,000)\end{tabular}  & \begin{tabular}[c]{@{}c@{}}\mcdb 610,500**\\(28,100)\end{tabular}    \\
\textbf{Female}         &                                                               & \begin{tabular}[c]{@{}c@{}}\mcdr ‐197,600***\\(62,000)\end{tabular} & \begin{tabular}[c]{@{}c@{}}\mcdr ‐269,000***\\(62,300)\end{tabular}  \\
\textbf{Serial}         &                                                               & \begin{tabular}[c]{@{}c@{}}\mcdr 777,000***\\(49,700)\end{tabular}  & \begin{tabular}[c]{@{}c@{}}\mcdr 745,900***\\(49,900)\end{tabular}   \\
\textbf{B.Sc.}             &                                                               &                                                               & \begin{tabular}[c]{@{}c@{}}50,490\\(52,900)\end{tabular}       \\
\textbf{M.Sc.}             &                                                               &                                                               & \begin{tabular}[c]{@{}c@{}}\mcdr 218,400***\\(54,500)\end{tabular}   \\
\textbf{MBA}            &                                                               &                                                               & \begin{tabular}[c]{@{}c@{}}\mcdb ‐203,300**\\(97,400)\end{tabular}   \\
\textbf{PhD}         &                                                               &                                                               & \begin{tabular}[c]{@{}c@{}}5675\\(96,000)\end{tabular}         \\
\textbf{Top university} &                                                               &                                                               & \begin{tabular}[c]{@{}c@{}}85,350\\(60,300)\end{tabular}       \\ 
\hline
\textbf{$R^{2}$}             & 0.032                                                         & 0.047                                                         & 0.065                                                          \\
\textbf{Country FE?}    & N                                                             & N                                                             & Y                                                             
\end{tabular}
\end{footnotesize}
\end{table}

\textbf{Gender}
The given results show that a startup with an higher share of female founders raises less money than companies with less female founders. Considering multiple other team variables such as \textit{percent serial founders present} or several fixed effects such as industry group and country, this finding remains evident. When comparing an all-male to a similar all-female startup, the latter raises, on average, around \$200,000 less in early-stage funding. Note that there is a linear relation based on the percentage of female founders, i.e. a startup with four founders would, on average, raise around \textit{\$50,000 to \$75,000 less per women involved}.
An open question is whether this gender effect is due to bias on the investor side or because women have set lower funding targets for their ventures than men. To this end, 
\citet{Ewens.2020} state that \textit{men participating in more risky businesses} than women as well as that \textit{male-led startups set higher targets}.

\textbf{Educational background}
Considering the coefficients for the educational variables, we observe that \textit{founder teams with MBA graduates raise smaller funding rounds}. This suggests that an MBA degree is, on average, less worth to investors than other academic degrees.
This result confirms the findings of 
\citet{Ewens.2020} showing that the probability of an investor interacting with or financing a startup is slightly lower when the CEO holds an MBA. 
Furthermore, a \textit{PhD in a founder team seem to have no impact on the amount raised}. Although higher education is expected to increase the intellectual capital of a startup, we assume that this effect is outweighed by a lack of work experience.
Moreover, \textit{graduating from the top 100 universities has no significant impact on the funding raised}. This suggests that founders from top universities neither receive better valuations compared to other entrepreneurs nor set higher fundraising targets. Hence, an analysis with a binary funding variable (\textit{received funding} -- \textit{no funding}) could be better suited. However, as outlined in Section \ref{analysis:limitations}, this information is not available on Crunchbase.

\begin{sidewaystable*}
\centering
\caption{Funding regression over time. %
Standard errors are reported in parentheses. Significance: \SignificanceColored.}
\label{tab:results-reg-years}
\resizebox{\textwidth}{!}{%
\begin{tabular}{lcccccccccc} 
\hline
\textbf{Variable}       & \textbf{2008}                                                     & \textbf{2009}                                                     & \textbf{2010}                                                       & \textbf{2011}                                                     & \textbf{2012}                                                     & \textbf{2013}                                                    & \textbf{2014}                                                     & \textbf{2015}                                                        & \textbf{2016}                                                     & \textbf{2017}                                                       \\ 
\hline
\textbf{Intercept}      & \begin{tabular}[c]{@{}c@{}}270,100\\(1,000,000)\end{tabular}      & \begin{tabular}[c]{@{}c@{}}\mcdg 1,505,000*\\(780,000)\end{tabular} & \begin{tabular}[c]{@{}c@{}}\mcdr 3,060,000***\\(780,000)\end{tabular} & \begin{tabular}[c]{@{}c@{}}302,200\\(226,000)\end{tabular}        & \begin{tabular}[c]{@{}c@{}}395,400\\(394,000)\end{tabular}        & \begin{tabular}[c]{@{}c@{}}237,400\\(51,460)\end{tabular}        & \begin{tabular}[c]{@{}c@{}}\mcdb 694,400**\\(325,000)\end{tabular}  & \begin{tabular}[c]{@{}c@{}}‐1,791,000\\1,130,000\end{tabular}        & \begin{tabular}[c]{@{}c@{}}\mcdb 686,700**\\(176,100)\end{tabular}  & \begin{tabular}[c]{@{}c@{}}\mcdb 1,162,000**\\(487,000)\end{tabular}  \\
\textbf{Female}         & \begin{tabular}[c]{@{}c@{}}‐663,500\\(556,000)\end{tabular}       & \begin{tabular}[c]{@{}c@{}}‐102,600\\(265,000)\end{tabular}       & \begin{tabular}[c]{@{}c@{}}\mcdr ‐610,200***\\(198,000)\end{tabular}  & \begin{tabular}[c]{@{}c@{}}\mcdb ‐259,600**\\(121,000)\end{tabular} & \begin{tabular}[c]{@{}c@{}}\mcdr ‐216,000***\\(74,100)\end{tabular} & \begin{tabular}[c]{@{}c@{}}\mcdg ‐189,200*\\(111,000)\end{tabular} & \begin{tabular}[c]{@{}c@{}}\mcdb ‐184,300**\\(90,900)\end{tabular}  & \begin{tabular}[c]{@{}c@{}}\mcdr ‐959,400***\\(271,000)\end{tabular}   & \begin{tabular}[c]{@{}c@{}}‐163,700\\(107,000)\end{tabular}       & \begin{tabular}[c]{@{}c@{}}‐43,720\\(123,000)\end{tabular}          \\
\textbf{Serial}         & \begin{tabular}[c]{@{}c@{}}‐348,600\\(246,000)\end{tabular}       & \begin{tabular}[c]{@{}c@{}}\mcdb 453,200**\\(206,000)\end{tabular}  & \begin{tabular}[c]{@{}c@{}}67,920\\(220,000)\end{tabular}           & \begin{tabular}[c]{@{}c@{}}2,288\\(78,800)\end{tabular}           & \begin{tabular}[c]{@{}c@{}}\mcdb ‐168,500**\\(76,400)\end{tabular}  & \begin{tabular}[c]{@{}c@{}}\mcdr 577,100***\\(93,900)\end{tabular} & \begin{tabular}[c]{@{}c@{}}‐56,110\\(66,100)\end{tabular}         & \begin{tabular}[c]{@{}c@{}}\mcdr 3,088,000***\\(223,000)\end{tabular}  & \begin{tabular}[c]{@{}c@{}}\mcdr 400,500***\\(87,100)\end{tabular}  & \begin{tabular}[c]{@{}c@{}}\mcdr 625,100***\\(101,000)\end{tabular}   \\
\textbf{Ba}             & \begin{tabular}[c]{@{}c@{}}‐185,600\\(200,000)\end{tabular}       & \begin{tabular}[c]{@{}c@{}}‐303,900\\(218,000)\end{tabular}       & \begin{tabular}[c]{@{}c@{}}\mcdr 661,100***\\(210,000)\end{tabular}   & \begin{tabular}[c]{@{}c@{}}\mcdr 322,300***\\(85,500)\end{tabular}  & \begin{tabular}[c]{@{}c@{}}34,660\\(73,100)\end{tabular}          & \begin{tabular}[c]{@{}c@{}}‐97,810\\(101,000)\end{tabular}       & \begin{tabular}[c]{@{}c@{}}‐103,100\\(73,300)\end{tabular}        & \begin{tabular}[c]{@{}c@{}}\mcdg 472,300*\\(264,000)\end{tabular}      & \begin{tabular}[c]{@{}c@{}}\mcdb ‐185,400**\\(80,800)\end{tabular}  & \begin{tabular}[c]{@{}c@{}}1358\\(119,000)\end{tabular}             \\
\textbf{Ma}             & \begin{tabular}[c]{@{}c@{}}\mcdr 132,400***\\(341,000)\end{tabular} & \begin{tabular}[c]{@{}c@{}}77,470\\(223,000)\end{tabular}         & \begin{tabular}[c]{@{}c@{}}83,320\\(154,000)\end{tabular}           & \begin{tabular}[c]{@{}c@{}}‐34,620\\(82,700)\end{tabular}         & \begin{tabular}[c]{@{}c@{}}\mcdr 229,000***\\(68,500)\end{tabular}  & \begin{tabular}[c]{@{}c@{}}118,300\\(98,500)\end{tabular}        & \begin{tabular}[c]{@{}c@{}}76,060\\(76,800)\end{tabular}          & \begin{tabular}[c]{@{}c@{}}\mcdb 547,700**\\(254,000)\end{tabular}     & \begin{tabular}[c]{@{}c@{}}71,770\\(93,400)\end{tabular}          & \begin{tabular}[c]{@{}c@{}}‐84,920\\(129,000)\end{tabular}          \\
\textbf{MBA}            & \begin{tabular}[c]{@{}c@{}}‐201,100\\(834,000)\end{tabular}       & \begin{tabular}[c]{@{}c@{}}220,800\\(448,000)\end{tabular}        & \begin{tabular}[c]{@{}c@{}}189,700\\(358,000)\end{tabular}          & \begin{tabular}[c]{@{}c@{}}198,700\\(138,000)\end{tabular}        & \begin{tabular}[c]{@{}c@{}}\mcdr 806,900***\\(129,000)\end{tabular} & \begin{tabular}[c]{@{}c@{}}‐16,550\\(181,000)\end{tabular}       & \begin{tabular}[c]{@{}c@{}}200,200\\(137,000)\end{tabular}        & \begin{tabular}[c]{@{}c@{}}\mcdr ‐2,036,000***\\(473,000)\end{tabular} & \begin{tabular}[c]{@{}c@{}}175,900\\(160,000)\end{tabular}        & \begin{tabular}[c]{@{}c@{}}\mcdb 532,900**\\(217,000)\end{tabular}    \\
\textbf{Doctor}         & \begin{tabular}[c]{@{}c@{}}621,900\\(814,000)\end{tabular}        & \begin{tabular}[c]{@{}c@{}}120,500\\(353,000)\end{tabular}        & \begin{tabular}[c]{@{}c@{}}116,800\\(248,000)\end{tabular}          & \begin{tabular}[c]{@{}c@{}}\mcdb 330,400**\\(153,000)\end{tabular}  & \begin{tabular}[c]{@{}c@{}}83,390\\(103,000)\end{tabular}         & \begin{tabular}[c]{@{}c@{}}216,300\\(181,000)\end{tabular}       & \begin{tabular}[c]{@{}c@{}}‐19,930\\(76,800)\end{tabular}         & \begin{tabular}[c]{@{}c@{}}\mcdg ‐773,900*\\(451,000)\end{tabular}     & \begin{tabular}[c]{@{}c@{}}\mcdr 426,200***\\(151,000)\end{tabular} & \begin{tabular}[c]{@{}c@{}}\mcdb 514,900**\\(243,000)\end{tabular}    \\
\textbf{Top university} & \begin{tabular}[c]{@{}c@{}}‐284,000\\(273,000)\end{tabular}       & \begin{tabular}[c]{@{}c@{}}\mcdb ‐626,900**\\(274,000)\end{tabular} & \begin{tabular}[c]{@{}c@{}}\mcdb ‐402,500**\\(60,300)\end{tabular}    & \begin{tabular}[c]{@{}c@{}}\mcdb ‐191,300**\\(94,800)\end{tabular}  & \begin{tabular}[c]{@{}c@{}}\mcdr ‐346,700***\\(77,400)\end{tabular} & \begin{tabular}[c]{@{}c@{}}‐135,500\\(99,500)\end{tabular}       & \begin{tabular}[c]{@{}c@{}}\mcdr ‐240,800***\\(80,400)\end{tabular} & \begin{tabular}[c]{@{}c@{}}\mcdr 1,802,000***\\(319,000)\end{tabular}  & \begin{tabular}[c]{@{}c@{}}\mcdr 493,400***\\(107,000)\end{tabular} & \begin{tabular}[c]{@{}c@{}}\mcdr ‐940,200***\\(160,000)\end{tabular}  \\ 
\hline
\textbf{Observations}   & 195                                                               & 327                                                               & 480                                                                 & 1,106                                                             & 1,620                                                             & 2,649                                                            & 2,712                                                             & 2,956                                                                & 2,331                                                             & 1,451                                                               \\
\textbf{$R^2$}             & 0.461                                                             & 0.430                                                             & 0.296                                                               & 0.197                                                             & 0.170                                                             & 0.096                                                            & 0.219                                                             & 0.169                                                                & 0.170                                                             & 0.207                                                              
\end{tabular}
}
\end{sidewaystable*}

\textbf{Serial founders} 
We observe that the presence of serial founders greatly increases the funding amount raised. Serial founders seem to have around three times more impact on the funding received than female founders. This effect shows that serial founders having more entrepreneurial experience exhibit a smaller risk towards investors and set higher funding target higher accordingly. Furthermore, we assume that serial founders have better network allowing them to raise funding for a new venture more easily.

\subsubsection{Impact over time}

Table \ref{tab:results-reg-years} presents the results from several OLS regressions done separately for each year in the examined period, with companies that received an early-stage funding that year. Similar to examined correlations coefficients over the year, many values in the 2008 and 2009 regression have low statistical significance. Unfortunately, many of the degree variables do not surpass a certain level of significance, although the results do have better p-values.

Considering the impact of female founders on the funding size over time, we see a positive trend of the coefficient. Note, the outlier in 2015 is significant lower, and go back up the following year. This indicates that although a gender effect is present, it is slowly decreasing over time. Overall, we cannot derive any detailed findings as many values are not statistical significant.

\subsubsection{Single Founders}

In our analysis, the majority of startups are founded by a single person. Therefore, it might also be interesting to analyze these ventures separately. Furthermore, we apply an OLS regression to compare female founders to male founders with respect to how each of their characteristics affect the amount raised. The female variable is omitted from the regression when analyzing each gender individually.
Table \ref{tab:results-reg-single} presents the results of the regression, which are further outlined in the following.

\textbf{Gender}
Female founders receive on average less funding than their male counterparts. This result corresponds to the previously reported findings. However, when comparing these numbers to Table \ref{tab:results-reg-all}, it seems that the gender of single founders has less impact on the funding amount than the presence of female founders in a team of founders has.

\begin{table}[tb]
\centering
    \caption{Funding regression by gender. %
    Standard errors are reported in parentheses. Significance: \SignificanceColored.}
    \label{tab:results-reg-single}
\begin{tabular}{lccc} 
\hline
\textbf{Variable}       & \textbf{Male Founder}                                                      & \textbf{Female Founder}                                           & \textbf{All}                                                       \\ 
\hline
\textbf{Intercept}      & \begin{tabular}[c]{@{}c@{}}\mcdr 910,600***\\(253,000)\end{tabular} & \begin{tabular}[c]{@{}c@{}}‐208,300\\(785,000)\end{tabular}       & \begin{tabular}[c]{@{}c@{}}\mcdr 888,000***\\(232,000)\end{tabular}  \\
\textbf{Female}         & —                                                                          & —                                                                 & \begin{tabular}[c]{@{}c@{}}\mcdb ‐91,450**\\(46,200)\end{tabular}    \\
\textbf{Serial}         & \begin{tabular}[c]{@{}c@{}}\mcdr 153,000***\\(41,700)\end{tabular}           & \begin{tabular}[c]{@{}c@{}}\mcdr 488,100***\\(185,000)\end{tabular} & \begin{tabular}[c]{@{}c@{}}\mcdr 144,600***\\(40,300)\end{tabular}   \\
\textbf{B.Sc.}             & \begin{tabular}[c]{@{}c@{}}‐40,180\\(42,400)\end{tabular}                  & \begin{tabular}[c]{@{}c@{}}‐44,590\\(119,000)\end{tabular}        & \begin{tabular}[c]{@{}c@{}}‐15,470\\(39,700)\end{tabular}          \\
\textbf{M.Sc.}             & \begin{tabular}[c]{@{}c@{}}29,190\\(44,300)\end{tabular}                   & \begin{tabular}[c]{@{}c@{}}‐116,200\\(122,000)\end{tabular}       & \begin{tabular}[c]{@{}c@{}}‐17,940\\(40,900)\end{tabular}          \\
\textbf{MBA}            & \begin{tabular}[c]{@{}c@{}}\mcdg 130,900*\\(75,900)\end{tabular}             & \begin{tabular}[c]{@{}c@{}}\mcdr 804,400***\\(210,000)\end{tabular} & \begin{tabular}[c]{@{}c@{}}\mcdr 223,600***\\(71,000)\end{tabular}   \\
\textbf{PhD}         & \begin{tabular}[c]{@{}c@{}}\mcdr 368,100***\\(75,700)\end{tabular}           & \begin{tabular}[c]{@{}c@{}}‐314,100\\(424,000)\end{tabular}       & \begin{tabular}[c]{@{}c@{}}\mcdr 342,200***\\(72,400)\end{tabular}   \\
\textbf{Top university} & \begin{tabular}[c]{@{}c@{}}‐60,660\\(49,600)\end{tabular}                  & \begin{tabular}[c]{@{}c@{}}\mcdr 878,400***\\(187,000)\end{tabular} & \begin{tabular}[c]{@{}c@{}}10,470\\(47,000)\end{tabular}           \\ 
\hline
\textbf{Observations}   & 4,877                                                                      & 447                                                               & 5,324                                                              \\
\textbf{$R^2$}             & 0.093                                                                      & 0.363                                                             & 0.090                                                             
\end{tabular}
\end{table}

\textbf{Serial Founders}
Being a serial founder seems to favor female founders way more than male founders. The difference between a female serial founder and a female first-time founder is three times as much as the difference between a male serial founder and a male first-time founder. This shows that gender bias in early-stage funding is less prevalent for serial founders with entrepreneurial experience.

\textbf{Educational Background}
We observe that higher education has a larger  effect for female single founders than for men. Having an MBA, women raise around \$800,000 more funding and men's funding amount increases only by \$130,000.

Furthermore, we find a positive effect for female single founders graduating from a top 100 university. In comparison, men neither raise greater nor smaller early-stage funding rounds when graduating from a top 100 university. For women, this effect is even greater than graduating with an MBA degree: The difference between graduating and not graduating from a top university amounts to about \$880,000 in money raised.
Additionally, it seems that women holding an MBA degree significantly offset the same coefficient in the regression done for all single-founder startups, despite their small portion in the sample.

In our analysis, only men profit from holding a PhD degree, receiving on average about \$370,000 more in early-stage funding money. Due to the small sample size, we cannot determine whether women raise higher or lower funding amounts holding a PhD.

\section{Main Findings}
\label{sec:main-findings}

The findings of our analysis can be summarized as follows: 
\begin{itemize} 
 \item Similar to related research, we found that women are underrepresented in the startup economy and startups with female founders raise on average smaller rounds than startups with an male-only founder team. Furthermore, startups with at least one female founder are relatively higher educated with a probability of having graduated from a top 100 university.

 \item %
We found a positive trend in the ratio of women-involved funded organizations over the years. This share increases very consistently from 6.62\% in 2008 up to 17\% in 2017. 
With the number of female founders increasing and, thus, likely joining the so-far male-dominant VCs and angel investors, we expect a development towards gender equality on both supply- and demand-side in the future.

 \item We found significant differences across European countries with respect to the participation of startups with female founders. For example, Italy and Bulgaria have a share of 21.5\% and 20\% respectively compared to Denmark's 8.47\% of female involved startups.

 \item Our correlation analysis shows a weakly negative correlation between the share of female founders and the funding size. Furthermore, the correlation coefficient is almost always negative, with some variances across industry groups. In this comparison, the hardware industry exhibits the largest gender gap. Moreover, gender effects are also observable across countries with some countries particularly favoring female founders, such as Sweden.

 \item Our regression analysis shows that the relative amount of female founders has a negative impact on the funding raised. Considering other founder characteristics and fixed effects across industries and countries, the negative impact persists. In contrast, serial founders seem to have a positive impact on the funding amount, three times as much than female founders have. Furthermore, we found that graduating from a top 100 university does not have an effect on the funding raised by a startup. The same applies for founders with a PhD.

 \item Considering single founder startups, we find that founder characteristics affect the funding raised differently based on the founders gender. Most significantly, female founders seem to profit three times as much from having already founded a startup. This shows that gender bias in early-stage funding is less prevalent for serial founders with entrepreneurial experience.
 Furthermore, we found that female founders raise larger rounds when graduating from a top 100 university or holding an MBA degree, whereas only male founders seem to benefit from a PhD.

 \item Comparing the results of both analysis methods, we found that regression analysis is better-suited providing results that are easier to interpret: Regression analysis allows us to directly compare the impact one variable on similar startups, whereas the correlation coefficients only indicate the direction of correlation. Furthermore, using regression we can account for fixed effects across industry groups and countries.
\end{itemize}
\section{Related Work}
\label{sec:related-work}

Research on gender effects in venture financing can both be divided by the source of the variables (demand- and supply-side) 
and the type of financing examined (e.g. bank loans, equity financing, crowdfunding). 
We will present related research on gender in financing based on this classification and relate their research question to our research. 
An overview of related work regarding the same type of variables is presented in Table \ref{tab:relatedwork}.

\subsection{Equity Financing via Venture Capital Firms}

\begin{sidewaystable*}
    \caption{Overview of related research regarding demand-side variables. *OLS (ordinary least squares) is assumed.}
    \label{tab:relatedwork}
\begin{scriptsize}
\hspace{0.4cm}
\hspace*{-1.1cm}\begin{tabular}{c@{\hskip 0.1in}l|l@{\hskip 0.1in}l|cl@{\hskip 0.02in}r|l@{\hskip 0.1in}l@{\hskip 0.1in}l} 
\hline
\multicolumn{2}{c|}{\begin{tabular}[c]{@{}c@{}}\textbf{Study} \\ \end{tabular}}        & \multicolumn{2}{c|}{\textbf{Data} }        & \multicolumn{3}{c|}{\textbf{Sample} }                                             & \multicolumn{3}{c}{\textbf{Regression} }                    \\
\textbf{Year}  & \textbf{Author}                                                       & \textbf{Origin}      & \textbf{Financing}  & \multicolumn{1}{l}{\textbf{Period} } & \textbf{Region}  & \textbf{\# Companies}  & \textbf{Model}  & \textbf{Variables}  & \textbf{Focus}      \\ 
\hline
2020           & \begin{tabular}[c]{@{}l@{}}We\\ \end{tabular}                      & \textbf{Crunchbase}  & Equity (VC)         & 2008-2017                            & Europe           & 4,900                   & OLS             & Team                & Funding Raised      \\
2019           & \begin{tabular}[c]{@{}l@{}}Ewens \& Townsend [8]\\ \end{tabular}     & AngelList            & Equity (Angel)      & 2010-2015                            & USA              & 17,780                  & OLS             & CEO                 & Investor Interest   \\
2019           & \begin{tabular}[c]{@{}l@{}}Raina [25]\\ \end{tabular}                 & \textbf{Crunchbase}  & Equity (VC)         & 2005-2013                            & USA              & 2,682                   & LOG             & Team                & Future Success      \\
2020           & \begin{tabular}[c]{@{}l@{}}Snellman \& Solal [31]\\ \end{tabular}    & \textbf{Crunchbase}  & Equity              & 2010-2018                            & USA              & 2,136                   & Cox             & Team                & Future Success      \\
2016           & \begin{tabular}[c]{@{}l@{}}Lins \& Lutz [20]\\ \end{tabular}         & KfW/ZEW              & Equity              & 2005-2009                            & Germany          & 6,793                   & OLS             & Team                & Funding Raised      \\
2019           & \begin{tabular}[c]{@{}l@{}}Lehto [19]\\ \end{tabular}                 & FVCA                 & Equity              & 2013-2018                            & Finland          & 474                     & -               & CEO                 & Funding Raised      \\
2018           & \begin{tabular}[c]{@{}l@{}}Brush et al. [5]\\ \end{tabular}           & Pitchbook            & Equity              & 2011-2013                            & USA              & 6,793                   & -               & CEO                 & Funding Raised      \\
2007           & \begin{tabular}[c]{@{}l@{}}Becker-Blease \& Sohl [3]\\ \end{tabular} & Survey               & Equity (Angel)      & 2000-2004                            & USA              & $\sim$40 angel portals    & OLS*            & CEO                 & Equity Surrendered  \\
2001           & \begin{tabular}[c]{@{}l@{}}Greene et al. [15]\\ \end{tabular}         & NVCA                 & Equity              & 1988-1998                            & USA              & 4,306                   & -               & CEO                 & Funding Raised      \\
2015           & Tinkler et al. [34]                                                   & Experiment           & Equity              & 2014                                 & USA              & 114 participants        & -               & -                   & Funding Raised      \\
2019           & \begin{tabular}[c]{@{}l@{}}Gafni et al. [10]\\ \end{tabular}          & Kickstarter          & Crowdfunding        & 2009-2012                            & USA              & 20,769                  & LOG             & Team                & Success             \\
2013           & \begin{tabular}[c]{@{}l@{}}Mollick [23]\\ \end{tabular}               & Kickstarter          & Crowdfunding        & 2009-2012                            & USA              & 2,101                   & LOG             & Team                & Success             \\
2019           & \begin{tabular}[c]{@{}l@{}}McGuire [22]\\ \end{tabular}               & \textbf{Crunchbase}  & Crowdfunding        & 2007-2015                            & USA              & 33,215                  & LPM             & Team                & Success             \\
2006           & \begin{tabular}[c]{@{}l@{}}Constantinidis et al. [6]\\ \end{tabular}  & Survey               & Loan                & 2001-2005                            & Belgium          & 288                     & -               & CEO                 & Loan Utilization    \\
2013           & \begin{tabular}[c]{@{}l@{}}Alesina et al. [1]\\ \end{tabular}         & Bank of Italy        & Loan                & \multicolumn{1}{l}{2004-2006}        & Italy            & 150,000                 & OLS             & CEO                 & Interest Rate       \\
2010           & \begin{tabular}[c]{@{}l@{}}Bellucci et al. [4]\\ \end{tabular}        & undisclosed          & Loan                & \multicolumn{1}{l}{2004-2006}        & Italy            & 7,800                   & LPM             & CEO                 & Interest Rate      
\end{tabular}
\end{scriptsize}
\end{sidewaystable*}

\subsubsection{Demand-side Analyses}
Early studies on gender effects in equity financing first revealed overall differences between genders via descriptive statistics:  
\citet{Greene.2001} found that -- although 36\% of US businesses were owned by women in 1996 -- of all companies funded by a VC firm from 1957-1998 only 2.4\% were identified as led by women. Identifying a slow but positive trend of women-led companies with a peak of 4.1\% in 1987, they also found that this gender gap varies between industries.

On top of that, a follow-up study by 
\citet{Brush.2018} showed that the gender gap in VC financed ventures further decreased: In the years from 2011 to 2013 the proportion of businesses receiving VC investments with a women on the executive team averages to 15\% -- more than three times the proportion reported for 1987. Additionally, they also looked at the amount of dollars invested. 
They conclude that female businesses receive greater sums on the average investment. \\
With regard to our research examining investments made between 2008 and 2017, we reflect on trends of women-led companies concerning the general amount of companies and their funding amount. In contrast to Greene et al. and most related work targeting US ventures (which is visible in Table \ref{tab:relatedwork}), we focus on the \textit{European} market.

Interestingly, Crunchbase was already in use for several gender bias analyses such as in 
\citet{Raina.2017} and 
\citet{Snellman.2020}. Table \ref{tab:relatedwork}, also gives an overview of datasets used in gender bias analyses.
Research as presented by 
\citet{Snellman.2020} also focuses on the investors gender. They investigated whether the gender of an early-stage investor affects future fundraising outcomes, by analyzing a sample that also originated from Crunchbase. In contrast to their research, we focus on the gender of the founders and do not include the investors gender. 

There is only little research on gender bias with regard to the European market as all aforementioned research focused on US ventures. By contrast, 
\citet{Lins.2016} examined gender differences of start-ups that utilises VC financing in Germany. Based on data from the KfW/ZEW Start-up Panel, they analyzed 3,317 "new ventures" founded between 2005 and 2009. Within their analysis, they excluded mixed-gender teams from their analysis in order to have more robust findings regarding gender differences. This aspect differs from our research, where mixed-gender founder teams are included in order to analyze the impact of the \textit{share} of female founders. 
Their research showed significant evidence of a gender gap in the use of VC financing and that higher levels of education were unable to bridge this gap. 
With regard to our research, we also analyze whether the educational status of founders influences the gender gap in funding amounts.

Another study analyzing gender effects outside the US market is provided by 
\citet{Lehto.2019}, which explored the existence of a gender gap in VC funding in Finland. In order to answer whether there are differences between women and female founders in VC funding and to what extent, he analyzed a dataset 
on Finnish startups funded between 2013 and 2018. 
Lehto discovers evidence for a gender gap in Finland and also finds that the Finnish gender gap in VC funding seems to be even bigger than what is reported for the US. This motivates our study to utilize international datasets and look more into gender effects across multiple countries.

\subsubsection{Supply Side Analyses}
\citet{Gompers.2014} focused on performance differences with regard to gender in the U.S. and investigated 3,437 VC partners and 26,087 investments. Collected data covered a period from 1975 up until 2003 originating from VentureSource \citep{VentureSource}, a large and well-known database regarding venture capital financing. Statistics showed an underperformance of female venture capitalists compared to male venture capitalists. 
On average, female investment performance was 15\% lower than male investment performance independent of firm or investment characteristics. In addition to that, female venture capitalists benefited less from more experienced colleagues. This effect was shown to be decreasing in larger, older companies having more female colleagues. 

Three years later, 
\citet{Gompers.2017} published further research on performance of venture capitalists, using data covering a period from 1990 to 2016 comprised of several types of information. They put into relation demographic and family background of general partners, VC firm hiring events as well as the deal and fund performances. 
They found partners having more daughters to hire more female venture capitalists.

\subsection{Equity Financing via Angel Investors}
\label{related-work:angel-investors}

\subsubsection{Demand-side Analyses}
As shown in Table \ref{tab:relatedwork}, there is fewer research on gender bias towards/shown by angel investors. For instance, 
\citet{Ewens.2020} examined gender effects in angel funding. They analyzed interactions between angel investors and 17,780 US companies listed on AngelList, 
an online platform for startups seeking early-stage funding as well as angel investors seeking investment opportunities. Their unique dataset allows them to not only analyze listed investments but also pre-funding interactions between startups and investors such as sharing a startups profile with another investor or requesting an introduction. In contrast to our research which takes all founders into account, they examined only the CEO. 
In contrast to our study and most other related research, they separated their analysis between male and female investors taking investors gender into account. They found that when comparing similar startups, female-led startups have a better chance of getting funded by a female investor. 
Their conclusion is that a gender bias is exhibited by angel investors favoring founders of their own gender.

Gender bias in angel funding regarding the US market was also in focus of 
\citet{BeckerBlease.2007}. In contrast to other studies, they also analyzed the rejection rate of funding showing that companies run by women were statistically as likely to be funded than companies run by men. Taking the founder perspective, they found that female businesses sought angel investors less often. On top of that, both genders tended to seek angel investors of their own gender which is often referred to as homophily. 

\subsubsection{Supply Side Analyses}
Harrison and Mason analyzed survey data collected around 2005 \citep{Harrison.2007}. They performed an exploratory study investigating gender effects of angel investors in the United Kingdom. Survey data comprised 40 angel investors with 21 women and 19 men. A quantitative analysis showed that gender did not affect the angels' investment decision, although female business angels were more likely to invest in women-owned businesses. %

\subsection{Other Financing Types}
Apart from equity financing, several researchers focused on gender bias in bank loans \citep{Constantinidis.2006,Alesina.2013,Bellucci.2010}. 
Furthermore, gender bias in crowdfunding has been studied \citep{McGuire.2019,gafni.2019,Mollick.2013}.

\section{Conclusion and Outlook}
\label{sec:conclusion}

In this paper, we studied whether the funding raised by a startup is affected by a founder's gender. If a gender effect was present, we were also interested in to know to what extent and how it compares to the influence of other founder-related characteristics. In order to answer these questions, we analyzed European startups listed on Crunchbase and their early-stage funding rounds from 2008-2017.

We found that gender effects and inequalities are still present. However, we also noticed a trend towards independence of gender related variables. Therefore, we encourage to keep studying gender effects and bias with respect to entrepreneurship and observe whether the trends found still persist in the future.

The framework in this paper was developed for understanding gender effects in the European startup scene. Similarly, this framework can as well be utilized for other regions such as for the emerging startup market in Asia and South America. Findings of such studies can then compare regions with each other in order to possibly reveal gender effects as the result of cultural differences between countries.

Furthermore, new types of funding a business emerged in recent years that have not yet been analyzed in detail. Some examples include equity crowdfunding and initial coin offerings in which corporate investors as the only funding source are replaced by the general public. 

\section*{Declarations}
\begin{sloppypar}
\textbf{Funding:} Not applicable.

\textbf{Conflicts of interest/Competing interests:} Not applicable.

\textbf{Availability of data and material:} The data used for our analysis was obtained by using the Linked Crunchbase API with a Crunchbase research license. Based on this license, we are not permitted to share the data. However, readers can retrace our analysis by using our code and by obtaining the data based on an own Crunchbase research license.

\textbf{Code availability:} Our code is available online at 
\url{https://github.com/michaelfaerber/crunchbias}.\end{sloppypar}



\bibliographystyle{elsarticle-num-names} 
\bibliography{bibliography}

\end{document}